Seismic Moment and Recurrence: Microstructural and mineralogical characterization of rocks in carbonate fault zones and their potential for luminescence and ESR dating

Evangelos Tsakalos[a,b,*], Maria Kazantzaki[a], Aiming Lin[b], Yannis Bassiakos[a], Eleni Filippaki[a], Nishiwaki Takafumi[b]

[a] *Laboratory of Luminescence dating, Institute of Nanoscience and Nanotechnology, National Centre for Scientific Research, N.C.S.R. "Demokritos", Athens, 153 10, Greece*
[b] *Department of Geophysics, Graduate School of Science, Kyoto University, Kyoto 606-8502, Japan*

***Corresponding Author:**

Evangelos Tsakalos, Laboratory of Luminescence dating, Institute of Nanoscience and Nanotechnology, National Centre for Scientific Research, N.C.S.R. "Demokritos", Aghia Paraskevi, 153 10, Athens, Greece. E-mail: e.tsakalos@inn.demokritos.gr; Tel.: +306974561999.

Abstract:

The important question of absolute dating of seismic phenomena has been the study of several researchers over the past few decades. The relevant research has concentrated on "energy traps" of minerals, such as quartz or feldspar, which may accumulate chronological information associated with tectonic deformations. However, the produced knowledge so far, is not sufficient to allow the absolute dating of faults. Today, Luminescence and Electron Spin Resonance (ESR) dating methods could be seen as offering high potential for dating past seismic deformed features on timescales ranging from some years to even several million years.

This preliminary study attempts to establish the potential of three different carbonate fault zones hosting fault mirror-like structures, to be used in luminescence and ESR dating, based on their microstructural, mineralogical and palaeo-maximum temperatures analysis. The results indicated that the collected samples can be considered datable fault-rock materials, since they contain suitable minerals (quartz) for luminescence and ESR dating, have experienced repeated cataclastic deformation and have been subject to various physical and chemical processes as well as pressure and temperature conditions.

# 1. Introduction

In the past years, new dating techniques and new methods of studying fault related rocks have motivated research into "direct fault dating" (e.g., Fantong et al. 2013; Bense et al. 2014; Shen et al. 2014). The establishment of the age of past earthquakes and knowing how often they have occurred in the past can provide the means to evaluate fault activity as well as the probability and severity of future earthquakes. In the field of direct dating of faults, several of the dating techniques which have been routinely applied to undeformed sedimentary deposits and archaeological materials, are still evolving in their application.

Today, luminescence and Electron Spin Resonance (ESR) dating methods could be seen as offering high potential for dating past seismic deformed features on timescales ranging from some years to even several million years (Rhodes et al. 2006; Sawakuchi et al. 2016; Tsakalos et al. 2016). Luminescence and ESR dating of fault movements are based on the assumption that the luminescence and ESR signals in fault-rock related features are reset during faulting as a result of the heat generated by friction and/or intense deformation (e.g., Nishimura and Horinouchi, 1989; Singhvi et al. 1994). Nevertheless, full resetting of the ESR and thermoluminescence signal of fault-rocks could not be established by some laboratory experiments (e.g., Toyoda et al. 2000) and measurements of fault-rocks (e.g., Fukuchi and Imai, 1998), while to the best of our knowledge, resetting of the optically stimulated luminescence signal has been partially established for experimentally sheared quartz gouge samples (Yang et al. 2018).

First, for dating of fault activity using these techniques, one needs to identify the geomorphic markers related (directly or indirectly) to the tectonic activities which are suitable for dating (features which have experienced frictional heat and/or shear stress and contain suitable minerals for dating). Consequently, for luminescence and ESR, an appropriate understanding of the microstructure and mineralogy of fault-rocks and related materials as well the host-rock, and the establishment of their potential to be used for dating is a requirement before any successful dating attempt.

The characterization of faults and their seismic-tectonic settings, along with their mechanical background and fault-rock formation processes, have been the subject of many studies and have been extensively reported in many published manuscripts and books (e.g., Hobbs et al. 1976; Scholz, 2002; Fossen, 2010, and references therein). Ductile and brittle deformation along faults most often leads to the development of fault cores and damage zones, comprising crushed rock fragments and fine-grained matrix (fault gouge, fault breccia and cataclasite), enveloped by the host-rock, that is frequently highly fractured (e.g., Sibson, 1977; Chester and Logan, 1986; Wallace and Morris, 1986; Caine et al. 1996). Fault-zone rocks show a diversity of geological structures and are classified principally by their textures and mechanical properties, e.g., if they are cohesive or foliated (e.g., Sibson, 1977).

Despite the plethora of data about fault-rocks and related materials, to date, microstructural and mineralogical characterisation of fault-rocks in carbonate fault zones is still in an relatively early stage. This comes in contrast to the enormous number of published studies on cataclastic fault zones in crystalline rocks and siliciclastic sediments (Di Toro et al. 2004; Wilson et al. 2005). As a result, the potential of carbonate fault-rocks to be used in geochronological studies could not be established.

The few existed studies however, have suggested that, in carbonate fault zones, brittle and cataclastic structures could be a common feature (e.g., Hadizadeh, 1994), with slickensides (also known as fault-mirrors) to be present. Fault slickensides could be found in limestone as well as dolostone fault zones and appear as polished surfaces (e.g., Jackson and McKenzie, 1999; Fondriest et al. 2013; Siman-Tov et al. 2013). Such features have been described to be made of a thin ultracataclasite layer, covering a rougher layer comprising calcite crystals and sitting on a relatively thick cataclasite layer (e.g., Siman-Tov et al. 2013). While the formation of these fault surfaces is not entirely understood, they are thought to represent seismic slip and strong frictional weakening (e.g., De Paola, 2013; Fondriest et al. 2013). Experimental studies have also managed to produce slickenside surfaces in simulated dolomite (Chen et al. 2013; Fondriest et al. 2013) and calcite (Boneh et al. 2013; Verberne et al. 2014; Siman-Tov et al. 2015) slip surfaces, indicating that such surfaces can be developed at seismic slip rates of ≥0.1 m/s and during dissipated frictional power density of 1-10 MW/m$^2$, which is analogous to that of natural earthquakes (Fondriest et al. 2013; Siman-Tov et al. 2015).

In view of the above, the presence of fault slickensides along natural carbonate faults could be interpreted as a geological marker of seismic slip, thus making fault slickensides and related formations good candidates for examining their suitability for direct dating of the past seismic activity of a particular fault.

Therefore, because of our limited knowledge concerning carbonate fault zones hosting mirror-like cataclastic structures, the aim of this study is to characterise, in terms of mineralogy and microstructure, such carbonate deformed features which are associated with past seismic events and determine their suitability for luminescence and ESR dating.

This paper is focused on microstructures and mineral compositions as well as palaeo-maximum temperatures of samples collected from three well developed carbonate fault slickenside formations and their surrounding host-rock in Italy and Greece, using optical microscopy, Scanning Electron Microscopy (SEM) backed by Energy Dispersive X-Ray Analysis (EDX) and X-Ray Diffraction (XRD) methods. The selection of the particular carbonate fault slickenside formations was based on evidence of faults activity (historical records of past fault raptures) as well as accessibility for sampling.

## 2. Study areas and geotectonic content

*2.1. Mattinata fault - Southern Italy*

The first study area was the Gargano Promontory in southern Italy (Fig. 1), a structural high, in the Adriatic foreland between the Apennines and the Dinarides-Albanides fold-and-thrust belts (Favali et al. 1993; Doglioni et al. 1994; Brankman and Aydin, 2004). It entails a thick succession of carbonates cut by an active fault system (the Gargano fault system). The shallow-water carbonate formation sits on dolomite and evaporitic deposits known as the Burano Formation (of Late Triassic age), which, overlay terrigenous paralic deposits of Permian-Early Triassic-age (Martinis and Pieri, 1964). Several major faults of diverse age and kinematics are evident in the region. Within this fault system, the Mattinata fault in the southern Gargano is the most profound fault (Ortolani and Pagliuca, 1987; Funiciello et al. 1988; Salvini et al. 1999; Brankman and Aydin, 2004). It has been associated with many large earthquakes. The 2016, Mw 6.2 earthquake in the Norcia region (where more than 250 people died) as well as the Mw ~5.0 that followed the 1893 mainshock (Mw ~5.4) in the Gargano Promontory, show that the Mattinata fault is active along its full length. The Mattinata fault has also caused a palaeoearthquake (Mw ~6.5) in 894 A.D (Piccardi, 2005). Many studies have dealt with this fault, however there still exist contradictory interpretations on its kinematics and tectonic history. In summary, the majority of studies suggest that at Mattinata, the main fault surface cuts through both recent slope deposits and bedrock units (Monte Saraceno Limestone) with its various types of kinematic indicators (e.g., slickensides and steps on striated surface) pointing out that the fault has moved right-laterally with a subordinate dip-slip (normal) component of motion (e.g., Todi et al. 2005).

At Mattinata, the main fault surface on the Monte Saraceno is, in places, well exposed in the form of a scarp and slickenside surfaces. During fieldwork, a well exposed fault slickenside on the main Mattinata fault was identified on the NE slope of the Monte Saraceno and three rectangular slickenside fault-rock samples were collected. Additionally, two host-rock samples were also collected for analysis.

**[Please insert Fig.1 here]**

*2.2. Arkitsa fault - Central Greece*

Our second studied-site is the spectacularly exposed fault slickensides of the Arkitsa fault zone, described by Jackson and McKenzie (1999), located on the southern side of the northern Gulf of Evia in central Greece (Fig. 2). The Arkitsa fault zone, of ~10 km in length, is a boundary between Late Triassic-Jurassic platform carbonates found in its footwall and Lower Pliocene to Quaternary sediments which develop in its hanging wall (Kokkalas et al. 2007). The exposed carbonate units sit on a Middle/Late Triassic volcano-sedimentary sequence (Pe-Piper and Panagos, 1989). A minimum 500-600m throw for the Arkitsa fault zone has been estimated, based on Neogene-Holocene sediment thickness on the hanging wall, the scarp height and topographic relief of the footwall block (Jones et al. 2009). Given that the activity of the faults of Evia rift zone initiated between 2-3 Ma (Ganas et al. 1998) a slip rate of 0.2-0.3 mm/yr may be proposed for the Arkitsa fault zone. The Arkitsa fault is best exposed in the distinctive slickenside of up to ~65 m in height (Kokkalas et al. 2007).

The majority of the normal faults in the region have been active during the Pleistocene, however the timeframe of their activity is not adequately established. Historical earthquakes in the area have been documented, but there is not a direct association with Arkitsa fault system or any other fault in the area with confidence. There is however, sufficient evidence to support the association of Arkitsa fault system with two rupture events in 1894, responsible for the development of ca. 30-km-long rupture at the surface (Ambraseys and Jackson, 1990). The magnitude has been estimated between Mw ~ 6.4 and Mw ~ 6.7 for the first shock and Mw ~6.9 and Mw ~7.2 for the second (Ambraseys and Jackson, 1990; Makropoulos and Kouskouna, 1994; Papazachos and Papazachou, 1997). However, the distinct geomorphological features of the area of Arkitsa with the back-tilted terraces on the hanging wall block may signify seismic activity along this fault during the Holocene (Papoulis et al.

2013). Three rectangular slickenside fault-rock samples were collected from the exposed carbonate fault slickenside of the Arkitsa fault. Additionally, one limestone host-rock sample was collected several meters away from the slickenside formation.

[Please insert Fig.2 here]

*2.3. Kera fault - Western Crete*

Our third studied-site is located in western Crete, Greece, a region sitting on the southern Aegean subduction zone, an area in Eastern Mediterranean which is considered as one of the most tectonically and seismically active (Kilias, 1994; Jolivet et al. 1996; ten Veen and Kleinspehn, 2003). In the peninsula of Spatha in western Crete the fault system of Malathiros-Rodopos (Fig. 3), is a complex neotectonic macrostructure, encompassing large grabens, horsts and wide fault zones, trending N-S and dipping to the west (Mountrakis et al. 2012). Several other smaller faults which run parallel to the main fault zone are also evident, as well as conjugate faults dipping eastwards. The fault system of Malathiros-Rodopos consists of three main segments; the Rodopos (18 km), Kera (4 km) and Malathira (9 km) segments (Mountrakis et al. 2012). The fault zone cuts Alpine basement formations as well as Miocene-Pliocene marl sedimentary formations. The activity of the fault segments is evident by the presence of a well-developed Pleistocene and younger scree (Mountrakis et al. 2012). Further, along all segments, epicenters of historic surface earthquakes (up to Mw ~7) are recorded (Stiros, 2001; Caputo et al. 2010). For these reasons, all segments are considered as active faults.

Our studied fault is the segment of Kera which is bent towards the NE, becoming a rather dextral transfer fault between the other two segments. Our investigations were conducted on a distinctively exposed fault slickenside of the Kera fault zone, where three rectangular slickenside fault-rock samples and one weakly consolidated breccia sample adjacent to the slickenside formation were collected. Additionally, one limestone host-rock sample was collected.

[Please insert Fig.3 here]

3. Materials and methods

In total, nine fault-rock samples were collected from the three carbonate fault zones and used for microstructural, mineralogical and palaeo-maximum temperature analysis. Host-rock samples were also collected for all three study areas for analysis. Initially, macroscopic observations and microstructural examinations allowed the characterization of the fault-rock samples collected. In this regard, the classification by Sibson (1977) and Fry (1984), revised by Lin (2008), was used to further characterise the samples. Among others, microstructural analysis is a vital tool to investigate the deformation mechanisms that produced the final structure of the collected fault-rocks, thus allowing assessing if such mechanisms could have affected their luminescence/ESR signal, and establishing their potential for direct dating.

Microstructural analysis was conducted through optical microscopy and scanning electron microscopy, backed by energy dispersive X-ray spectroscopy. To observe and characterize deformation microstructures, polished petrographic thin sections of the fault-rock and host-rock samples were prepared and studied under transmitted as well as reflected light. Photographs of the thin sections were made by using the Microscope NIKON LV100POL and the coupled digital camera Nikon DS-Ri2. Using SEM/EDX, samples were scanned initially to map elements and determine distinctive separations, boundaries, and morphologies, but also grain shape and grain size characteristics and microstructures associated with the conditions prevailing during deformation. A 20 mm in diameter polished thin section for each sample was carbon-coated prior to examination. Typically, operating conditions used in the SEM analyses were: voltage 20 kV, magnifications from ~50 to 10,000X and working distances from 10 to 12 mm. Backscattered electron imaging (BSE) was used, as the grey level can give information about chemical contrast of the different features. Analytical EDX settings were 20 kV beam voltage and count time of 80-100 seconds (peaks). EDX analysis provided information on elements having concentrations at roughly 5% level or greater. The samples were examined in a JEOL JCM-6000 Plus SEM fitted with a semi-quantitative Energy-Dispersive X-ray (EDX) analyser, at the Department of Geophysics of Kyoto University, Japan.

Mineralogical analysis of fault-rocks can give information about the presence of minerals suitable for luminescence and ESR dating, while it also provides a significant contribution in assessing the conditions that prevailed during fault-rock formation, including palaeotemperature regime (e.g., Wu et al. 1975; Evans and Chester, 1995; Morrow et al. 2001; Kitagawa et al. 2007; Schleicher et al. 2009). In this regard, XRD analysis on both the bulk samples and their clay size (<2 μm) fraction was employed using a representative part of each sample. For bulk XRD analysis, preparation of the samples included air drying and homogenization through grinding in an agate mortar. For clay mineral XRD analysis, carbonates were removed using acetic acid, starting from very low acid concentration of 0.1 N to 1 N, depending on the carbonate content of each sample. This procedure secures carbonates removal and leaves unaffected the randomly interstratified illite-montmorillonite, chlorite, illite, or kaolinite (Ostrom, 1961). To avoid possible mixing of detrital with authigenic clays, the samples were not powdered but only gently disaggregated. Following carbonate removal, clay was deflocculated using consecutive washing with distilled water. The <2μm fraction was isolated by centrifugation and smeared onto zero-background holders for XRD analysis. This approach has the advantages of using only a few mg of sample while the zero-background holders, made of Si single crystal, give no background, allowing very minor components to be detected. The disadvantage is that the holders cannot withstand heating for further characterization of the clay mineralogy. The thickness of the prepared mounts exceeded 50 μm, which is a requirement for semi-quantitative determination of the clay mineral content with confidence (e.g., 'infinite thickness' of Moore and Reynolds, 1997).

The analyses were made on a SIEMENS D-500 automated X-ray diffractometer with a Cu-K$_\alpha$ anticathode at the Institute of Nanoscience and Nanotechnology, National Centre for Scientific Research, "Demokritos", Greece.

The bulk samples were scanned at a step of 0.03°/2sec from 2° to 60°, while the air-dried clay fraction samples were scanned from 2° to 70° with a step of 0.03°/12sec. The relative abundances of different minerals were assessed and estimated from integrated peak areas.

XRD analysis of the clay minerals viz. illite and chlorite was of particular interest in this study, since it has been demonstrated that clay mineralogical data on illite and chlorite can provide a reasonable estimate on the thermal excursion (and thus on resetting of the luminescence/ESR signal) during faulting (e.g., Singhvi et al. 1994; Banerjee, 1996). The illite analysis was based on the methodology suggested by Kübler (1967) which was subsequently adopted by Johnson et al. (1985), where the peak width at half-height of the 10 Å illite peak is used to differentiate between the diagenetic zone (<200 °C), the anchizone (200-370 °C) and the epizone (> 370 °C). On elevated temperatures, illite releases interlayer water and takes up potassium. A rearrangement of ions takes place as potassium ions attached into the interlayer parts, which is depicted by a sharp XRD peak at 10 Å. Weaver (1960) was the first to use this observation to assess low-grade metamorphism, where late Weber (1972) developed an illite Crystallinity Index (known as Hb$_{rel}$) expressed as:

Hb$_{rel}$ = [Hb (10 Å)$_{illite}$ / Hb (4 Å)$_{quartz}$] x 100,

(where Hb(001) illite and Hb(100) quartz are half-peak-widths of the (001) and (100) reflections of illite and quartz respectively).

Quartz, being generally unaffected by heating was used as a standard for interlaboratory comparisons of the $Hb_{rel}$ values. We used the $Hb_{rel}$ upper and lower limits of the anchizone proposed by Johnson et al. (1985), being 278 and 149 respectively.

Regarding chlorite, we followed the approach of Banerjee (1999). In the case of chlorite, it has been demonstrated that on heating, the XRD patterns of chlorite show a transformation in the peak-areas ratio (Ch peak-ratio) at 7.1 Å and 14.2 Å peaks. In particular, it has been demonstrated that temperatures between 0-250 °C and 250-400 °C produce peak-ratios of 2.0 ± 0.1 and 1.6 ± 0.2 respectively, while heating at 400-500 °C leads to an almost zero ratio, as minerals of the chlorite facies become chlorite schists and sericite at these temperatures (Dunoyer de Segonzac, 1970). For this study, the assessment of palaeotemperatures using the minerals illite and chlorite was performed on fault-rock and host-rock samples. Host-rock samples were included to examine the robustness of this approach, since it is expected that host-rock samples were not subject to any high temperature and thus they should not show any heating signatures.

We considered it important to use both illite and chlorite minerals in our palaeotemperature assessments, as it is usually problematic to differentiate illite from muscovite and chlorite from kaolinite with a reasonable confidence level. In addition, many studies using Ch peak-ratio have suggested a poor correlation with $Hb_{rel}$ values (Dalla Torre et al. 1996; Wang et al. 1996; among others), which indicate that Ch peak-ratio may be less reliable for palaeotemperature examinations. Reasons for this lack of reliability include the behaviour of chlorite, which tends to retain more dislocations than illite in anchizone and epizone, something that contributes to polygonization and segmentation of the crystals leading to the formation of subgrains and reduction of the total crystallite size (Merriman et al. 1990).

Thus, it should be noted here, that the results produced through the approach adopted in this study, must be considered with caution. Several other crystallinity indices exist and have been proposed for both illite and chlorite that can provide information on temperature excursion (e.g., Weaver, 1960; Dunoyer de Segonzac et al. 1968; Kübler, 1968; Árkai, 1991). Their employment however, necessitates calibration of the indices data, through the use of set of standards in the form of rock fragments provided by various laboratories, which require full preparation by the user. It is our purpose to explore the use of these indices in the near future, to provide additional constrain on our current results.

Following XRD analysis, SEM examination was performed on the same whole rock samples as well as clay size separates analysed by XRD. Regarding clay size fraction, for each sample, a flake of the aggregated clay was dispersed in distilled water and deposited by pipette onto a carbon-coated tape. A further carbon-coat was applied to minimise charging of the clay particles. The analyses were performed to provide complimentary information on samples mineralogy as well as mineral elemental concentrations through EDX. Single clay minerals are difficult to be differentiated by means of SEM/EDX due to variable contents of volatile components and interchangeable cations as well as the presence of mixed layer clay phases (e.g., Blanco et al. 2003). However, certain patterns in elemental composition may be observed which could enable distinction between clay minerals. In this study, mineral phases were examined based on minerals distinctive morphologies and calculating their stoichiometric ratios from atomic % of constituent elements, acquired by the quantitative EDX analysis with an acquisition time of 80-100 sec, and comparing them with known stoichiometric ratios obtained from published mineral databases (Anthony et al. 2009; Barthelmy, 2010).

## 4. Results and discussion

### 4. 1. Microstructural analysis

Throughout this section we refer to both the slip surface and slip zone when describing fault-rock-samples collected from the slickensides formations. The slip surface is the fault "plane" itself, while the slip zone refers to the formation that lies below the slip surface. The slip surface and the slip zone together, hold the bulk of co-seismic displacement during individual rupture events (Sibson, 2003; Smith et al. 2011). Microstructural analysis was performed on thin sections prepared for each of the fault-rock and host-rock samples. For the fault-rock slickenside samples, thin sections were cut perpendicular to the slip surface.

*4.1.1. Mattinata fault - Southern Italy*

The slickenside is characterised by a slip zone of a total thickness of up to five centimetres and appears as an uninterrupted stratum of cataclasite, beneath the principal slip surface (Fig. 4a). The slip surface contains a 40-50 μm thick distinctive brownish in colour slip layer which makes a sharp boundary between the slip surface and the slip zone cataclasite. Clasts belonging to the slip zone appear truncated when in the contact with the slip layer above. Two sub-layers are identified within the slip surface (Fig. 4b) running parallel to the slip surface; however in places they truncate one another (Fig. 4b). Sub-layer 1 (closer to the outer part) appears as a light brown ultracataclasite of approximately 300 μm in thickness with a very fine calcite matrix containing angular to sub-rounded clasts. Sub-layer 2 is thicker (~500 μm) than sub-layer 1 and appears as a dark brown/grey ultracataclasite with a fine calcite matrix containing angular to sub-rounded clasts. In places, sub-layer 2 contains clasts of fossiliferous limestone (most probably derived from the host-rock). Both sub-layers are dominated by the presence of calcite vein networks and veinlets of various thickness and geometry (Fig. 4c) with calcite crystals showing a blocky morphology and strong twinning (Fig. 4d). The cataclasite (slip zone) beneath the slip surface is 4 to 5 cm thick and encloses angular to sub-rounded poorly sorted clasts in a fine matrix (Fig. 4a). Fractures are a common feature and appear to break apart angular clasts. Individual calcite veins and vein networks are also evident throughout the slip zone and they are commonly broken apart by fractures. The extensive development of calcite veins indicate fluid infiltration and healing of fractures and may be attributed to circulation of a late-stage fluid phase, while the fact that some veins appear to have been broken apart may represent a later co-seismic event, post-dating that of the initial cataclastic deformation.

**[Please insert Fig.4 here]**

SEM-EDX analyses of the slip zone cataclasite and slip surface ultracataclasites illustrate that the matrix is mainly made of calcite grains, minor quartz, lesser amounts of clay minerals (illite, chlorite and kaolinite) and iron oxides. The minor clay minerals illite and kaolinite appear in the form of very small patches and aggregates (Fig 5a). Quartz appears with a high level of corrosion, overgrowths in placed corroded, and dissolution (etch) pits of varying density, size and shape (Fig 5b). Quartz overgrowth corrosion seems to be associated with kaolinite and illite presence (Fig. 5c), however in some cases kaolinite formation appears to be synchronous with overgrowth formation. SEM/EDX observations showed that illite might be formed first while kaolinite started replacing illite in a later stage. This observation suggests that a fundamental change in fluid composition may have taken place, since high $a_{K+}/a_{H+}$ (potassium/hydrogen ion activity) favours the growth of illite and low $a_{K+}/a_{H+}$ the formation of kaolinite (e.g., Yates and Rosenberg, 1997). Further, the observation that in some cases, kaolinite formed at the same time as the quartz overgrowths, suggests high $a_{Si4+}$ and low $a_{K+}/a_{H+}$ (e.g., Offler et al. 2009). The presence of etch pits in quartz overgrowths along with corroded quartz grains signify that conditions changed after the development of these minerals. These observations signify that the fluid was not in equilibrium with quartz and that $a_{Si4+}$ should have been low in the fluid (e.g., Offler et al. 2009).

Optical observations and EDX analysis of the host-rock samples, collected several meters far from the slickenside, revealed that they are a typical limestone with a matrix consisting of skeletal fragments of marine origin, grained calcite (micrite), iron oxides, minor amounts of clay minerals and some slightly elongate quartz grains (Fig. 5d,e).

[Please insert Fig.5 here]

The SEM-EDX investigation on the clay size fraction of the slickenside fault-rock samples revealed illite forming irregular flake-like platelets oriented parallel to each other (Fig 6a), while in places, illite also occurs in a rod-like form. Analysis on illite through EDX revealed the presence of the major elements: O, Si, Al, and K, with a minor amount of Mg, Ca, and Fe. Individual crystals of smectite cannot be determined in the SEM. Further, illite-smectite mixed-layer phases identification based only on EDX spectra analysis and morphology is very difficult because their composition and morphology are highly variable. Nevertheless, a distinct filamentous morphology (Fig. 6b) backed by EDX analysis (O, Si, Al, K, Ca, Mg, Fe and Cl) indicates a mixed layer of illite-smectite in the fault-rock samples. The elements Cl and Fe may be considered contaminants from adjacent minerals. Chlorite is also present, forming rosettes of individual platelet crystals (Fig. 6c). EDX analysis produced a characteristic spectrum for iron-rich chlorite (O, Si, Al, Mg, and Fe). Additionally, kaolinite was observed in our samples in the form of face-to-face stacks of pseudohexagonal plates or books (Fig. 6d), while EDX analysis yielding approximately equal peak heights of Si and Al confirmed its presence.

[Please insert Fig.6 here]

*4.1.2. Arkitsa fault - Central Greece*

The fault-rock samples collected from the slickenside formation of Arkitsa typically consists of a two to four centimetres thick cataclastic slip zone beneath a well defined ultracataclastic slip surface and a very thin slip layer which makes a boundary between the slip zone and the slip surface (Fig. 7a). The slip zone contains highly angular to sub-rounded clasts of limestone and minor altered dolomite and quartz grains in a fine-grained, iron oxide impregnated, intergranular matrix (Fig. 7a). Truncated clasts are evident when they touch the slip layer. In places, the matrix is highly variable in grain size, from visible fragments around 0.1 mm to submicroscopic. Towards the slip surface, the cataclastic texture passes to an ultracataclastic containing clasts of limestone-dolomite surrounded by a finer-grained matrix (Fig. 7b). Under the optical microscope, the larger calcite and quartz grains show undulose extinction with intragranular cracks which are frequently filled with calcite or matrix material. Fragments of dolomite (<200 μm in diameter), having minor undulose extinction can be seen dispersed into the cataclasite (Fig. 7c).

A distinctive network of discontinuous veins is evident in both the slip zone and the slip surface, signifying extensive fracturing (Fig. 7d). Earlier developed veins are in places fractured and asymmetric, whereas later developed veins are undeformed. Calcite and quartz within veins show irregular grain boundaries and range from being profoundly twinned and sheared to almost strain-free (Fig. 7e). Boundaries between quartz grains and calcite are serrated and bulged (recrystallized) (Fig. 7e). Serrated grain boundaries may indicate the effect of either pressure solution or of strain induced migration of grain boundaries (Spry, 1969) while the bulges could be seen as the initial step of recrystallization (White, 1977). The observed recrystallization could also be attributed to temperature, as it has been shown that strain induced migration starts at about 275 ⁰C and recrystallization at about 290 ⁰C (e.g., Voll, 1976).

Further, in places, very small, strain free new quartz grains with asymmetrical, lobate grain boundaries develop due to recrystallization within the most intensely deformed grains (Fig. 7f). The observation supports a static recrystallization regime rather than a dynamic grain boundary recrystallization process (e.g., Jessell, 1986; Urai et al. 1986; Poirier and Guillope, 1979).

**[Please insert Fig.7 here]**

The photomicrographs obtained on polished thin-sections of a host-rock (limestone) sample collected several meters away from the slickenside of Arkitsa are given in Fig. 8. The matrix is calcilutite, and consists of minute recrystallized carbonate grains and ooids and some minor quartz grains, in a finely divided, slightly iron-stained, calcite base (Fig. 8a). The matrix shows an extensive network of veins and veinlets which are filled with sparry calcite grains (Fig. 8b).

[Please insert Fig.8 here]

Further SEM-EDX analysis of both the slip zone and the slip surface confirmed the observations made under the optical microscope. The matrix is mainly composed by calcite grains, minor quartz, iron oxides and lesser amounts of clay minerals (illite/smectite and kaolinite) and dolomite. Minor clay minerals illite/smectite and kaolinite form small grain size patches (Fig. 9a). Quartz appears with a high level of corrosion (Fig. 9b), in places associated with illite and kaolinite minerals. Quartz overgrowths can also be observed but are not very common.

[Please insert Fig.9 here]

Backscattered electron microscopy images combined with EDX analyses of the fault-rock samples revealed a variety of illite, illite-smectite, chlorite and kaolinite of different shapes, and sizes (Fig. 10). A variety of illite crystal shapes; reasonably large subhedral to euhedral, along with smaller illite particles having distinctive irregular edges can be observed (Fig. 10a). EDX analysis yielded the major elements O, Si, Al, and K, with a minor amount of Mg. Illite-smectite mixed-layer may also be presented, but this is difficult to be identified by SEM with confidence, because of its highly variable composition and morphology. Nonetheless, EDX analysis of a clay aggregate showing filamentous morphology (Fig. 10b) revealed the major elements O, Si, Al, K, Ca, Mg, and Fe, which may be interpreted as illite-smectite mixed layer minerals. Chlorite is also evident in our fault-rock samples from Arkitsa but it is not as common as illite and illite-smectite mixed layer, with EDX analysis of platelets arranged in an almost a rosette pattern (Fig. 10c) yielding the major elements O, Si, Al, Mg, and Fe. Kaolinite can also be observed in the form of face-to-face stacks of pseudohexagonal plates (Fig. 10d) with EDX analysis producing nearly equal peak heights of Si and Al.

[Please insert Fig.10 here]

*4.1.3. Kera fault - Western Crete*

As in Mattinata and Arkitsa, the fault-rock samples from Kera slickenside are characterised by a slip zone of cataclastic fabric (2-4 cm thick), supported by randomly oriented angular to subrounded clasts and in places cross-cut by fracture networks and stylolite dissolution surfaces (Fig. 11a,b). The slip zone sits beneath a sharp slip layer and a well-developed ultracataclastic slip surface (Fig. 11a). Clasts in the slip zone of Kera slickenside formation also appear truncated when in the contact with the slip layer above. The presence of stylolites at high angles suggests that pressure solution in a tectonic manner was also an important deformation mechanism (e.g., Railsback and Andrews, 1995). Veins also develop throughout the slip zone cataclasite, in places cross cut by open fractures (Fig 11c). Within the veins, calcite usually shows a heterogeneous grain size distribution, undulose extinction and moderate twinning (Fig. 11d). The ultracataclasite slip surface appeared to have experienced more intense grain comminution than the slip zone cataclasite, forming discontinuous patches and plagues above the slip zone due to fault scarp erosion.

Both the slip zone cataclasite and slip surface ultracataclasite show a matrix of variably sized fragments of angular to rounded grey-brown to dark-grey calcite, quartz, dolomite, calcite aggregates, quartz aggregates, in places showing iron oxide impregnation (Fig. 11e). Clasts in the matrix also have intragranular and intergranular fractures which are commonly filled with fine calcite. Under optical microscope the boundaries between the matrix and the calcite grains appear highly variable, irregular and partially serrated, although observations are limited by the resolution of the optical microscope (Fig. 11f). Only a few calcite grains are affected by subgrain development.

**[Please insert Fig.11 here]**

The host-rock (Fig. 12 a,b) is a limestone made of almost exclusively by abraded/rounded carbonate grains (fossils/skeletal grains, quite well sorted) and calcite cement as well as minor dolomite and quartz. The minor dolomite is rhombohedral in form, while dolomite lenses have sutured boundaries (Fig. 12b). Some highly corroded calcite veins are also evident in the hot-rock (Fig. 12a).

A fault-breccia sample collected from an outcrop adjacent to the slickenside was also examined. This sample is a medium-crushed inconhesive breccia that contains weakly disaggregated structures and clast sizes that are typically less than 400 μm, in a very finely granulated matrix (Fig. 12c). It appears light brown in hand specimen and brown in thin sections. It is principally composed of carbonates and quartz, opaque minerals (oxides), and more rarely muscovite (detrital) (Fig. 12c). Quartz and clay minerals with a high birefringence (phylosilicates) compose the matrix. The weakly disaggregated fault-breccia particles usually have a chaotic structure and lack a clast shape preferred orientation. Precipitation of Fe phases is commonly visible on the margins of the angular clasts. A small number of antithetic microfractures in clasts (mainly in calcites) is also a feature (Fig. 12d), most probably indicating localised deformation (e.g., Desbois et al. 2017). SEM/EDX observations of the breccia sample showed that, in places muscovite has been replaced by fine-grained aggregates of illite and kaolinite (Fig. 12e). The presence of kaolinite along the basal cleavages of muscovite has been noted by a number of authors (e.g., Johnson et al. 1992; Milliken, 2003, and references therein). It indicates that the basal cleavage surfaces of muscovite have been exploited by fluids responsible for the formation of kaolinite, more probably in a low temperature diagenetic regime, where dissolving muscovite grains gradually loses K, Mg and Fe from interlayer and octahedral sites (Bjorlykke et al. 1979; Burley, 1984; Kantorowicz, 1984).

[Please insert Fig.12 here]

Further, quartz in both slickenside fault-rock samples and breccia sample exhibits overgrowths and dissolution (etch) pits of varying density, size and shape (Fig. 13). The pits occur on the surface of quartz grains and on quartz overgrowths. They more commonly have triangular, cusp and rectangular shapes (Fig. 13a,b). The overgrowths appear as euhedral crystals attached to individual quartz grains and their development is connected with neocrystalline kaolinite and illite in the slickenside fault-rock samples. Overgrowths are better developed in the slickenside fault-rock samples than in the breccia sample. The triangular-shaped etch pits and the observed overgrowths developed in quartz have been noted in many studies (e.g., Blum et al. 1990; Tsakalos et al. 2015; Tsakalos, 2016) and have also been produced experimentally on quartz by Brantley et al. (1986) and Joshi and Kotru, (1969). According to Brantley et al. (1986 and references therein), the formation of such etch pits along with their number and deepness, are related of the concentration of Si in the fluid phase and the time of the etching period. Dislocation tangles commonly lead to preferential dissolution and cause diffusive mass transfer, which subsequently produce etch pits (Brantley et al. 1986; Blum et al. 1990).

Additionally, the concentration of Si in the fluid phase which is required for the development of the quartz overgrowths presumably has resulted from the breakdown of detrital phases and as well as pressure solution. The sutured contacts between quartz grains are a clear indication that pressure solution has been taken place. However, to explain the extensive formation of overgrowths observed in the slickenside fault-rock samples, an external supply of Si is more likely. Influxes of fluid through the fault zones are very possible, transporting amounts of Si which were required for the development of the extensive overgrowths observed. We consider silica precipitation as quartz overgrowth on pre-existing surfaces, governed by the kinetic equation of quartz overgrowth (Rimstidt and Barnes, 1980).

Conchoidal fractures (curved, shell-like breakage patterns) can also be observed on quartz grains in both slickenside fault-rock samples and the breccia sample (Fig. 13c,d) from Kera. The presence of conchoidal fractures is interpreted as the product of powerful impact or pressure on the grain surface (e.g., Vos et al. 2014; Tsakalos, 2016). As pressure is transmitted through quartz crystal lattice, it generates an uneven appearance on quartz surface (Kragelskii, 1965; Margolis and Krinsley, 1974). Further, some small holes (< 1 μm) are commonly seen on quartz fractured surfaces. They are thought to be the result of mineral inclusions (Le Ribault, 1977), which can cause weakening of the quartz crystal lattice, leading to the formation of the fracture-hole.

[Please insert Fig.13 here]

SEM observations of the clay size fraction of slickenside fault-rock samples revealed the presence of illite as filamentous aggregates and commonly as lath-like crystals (Fig. 14a). EDX scans detected the major elements: O, Si, Al, and K, with a minor amount of Mg, and Fe. The observation of a thin webby mineral morphology which is a common crystal habit of smectite was also detected and was interpreted as smectite or illite/ smectite mix layer (Fig. 14b). However, this interpretation should be considered with caution, since this characteristic is distinctive but not unique to smectite. Further, analysis of the EDX spectra may be problematic due to electron beam penetration through the clay and into the underlying substrate, producing a composite EDX spectrum. Here, EDX spectrum gave the major elements of O, Si, Al, Ca, Mg, Fe, and K. The peak height of Al compared to Si is much higher than the expected for illite-smectite. SEM/EDX analysis showed underlying kaolinite, which possibly contributed additional Al to the EDX analysis.

Chlorite was also identified in the fault-rock samples from Kera fault, which appears in the form of dispersed platelets usually arranged perpendicular to one another (Fig. 14c). EDX analysis yielded a typical spectrum for iron-rich chlorite, containing the major elements O, Si,

Al, Mg, and Fe. Finally, kaolinite was also observed in the form of plates and book-like structures (Fig. 14d), with a typical kaolinite EDX spectrum.

[Please insert Fig.14 here]

*4.2. Mineralogical analysis*

*4.2.1. Mattinata fault - Southern Italy*

XRD bulk examination of the fault-rock and host-rock samples from Mattinata revealed that they are dominated by calcite, quartz and very low proportions of clay minerals and goethite. The results are presented in Table 1 in terms of minerals' relative abundances (minerals present in abundances greater than c. 3%). Clay minerals through bulk XRD analysis showed very weak peaks, making difficult to distinguish the different phylosilicates. Nevertheless, peaks patterns indicate the presence of minor illite/muscovite and chlorite/kaolinite. The low illite/muscovite content may indicate a relatively low permeability as well as low intensity of fluid-rock interaction (e.g., Surma and Geraud, 2003).

XRD analysis of clay size (<2μm) separates shows a composition of predominately quartz, minor illite/muscovite and lesser chlorite/kaolinite. Microcline was found in one fault-rock sample (MAT7). The characteristic peak for 001 kaolinite at c. 0.7 Å overlaps with the 002 chlorite peak; thus we here refer to both minerals. Based on the peak shape, SEM/EDX examination as well as the occurrence of a weak kaolinite 002-peak at 0.4 Å, both minerals most likely are present in all samples (see also Biscaye 1964). Further, the 001 illite peak overlaps with 001 muscovite peak, thus we refer to both illite/muscovite minerals.

[Please insert Table 1 here]

The presence of kaolinite and quartz in the fault-rock samples suggest that maximum temperature never exceeded approximately 300 ºC (Winkler 1976, p. 204). Further, kaolinite may suggest a hydrothermal episode with more acidic conditions and leaching cations, or it may reflect a late alteration by low-temperature surface water (Ratcliffe and Burton, 1988).

Several studies have shown similar mineralogy as the Mattinata fault zone with smectite, illite chlorite and kaolinite being the main clay minerals (e.g., Hashimoto et al. 2007; Isaacs et al. 2007; Sone et al. 2007). In case of Mattinata however, smectite was not observed (based on SEM/EDX examination and the relatively narrow/steep appearance of the illite XRD peak) in the fault-rock samples, but a low content of smectite in an illite-smectite mixed layer (based on SEM observations and EDX analysis) is more likely apparent. Hashimoto et al. (2007) proposed that smectite depletion within a fault zone may take place due to dehydration as a result of shear heating, leading to a mixed layer of illite-smectite, with temperature being the main factor influencing this reaction. Alternatively, this may be due to faulting itself which could provide the energy required to surpass the kinetic barrier and transform smectite into illite (Vrolijk and van der Pluijm, 1999; Isaacs et al. 2007).

Using $Hb_{rel}$ values, the fault-rock samples of Mattinata fell within the range of 200-370 ºC (anchizone), while the Ch peak-ratios for the same samples indicated a palaeotemperature in the range of 250 to >400 ° C (Table 2). Both $Hb_{rel}$ values and Ch peak-ratios indicate that besides the fault-rock samples, the host-rock was also heated to a high temperature. This result is confusing, since the host-rock is a limestone and should not present any heating signatures. This discrepancy may arise due to the fact that, minerals illite and chlorite were derived from a common metamorphic provenance or due to complications associated with the possible overlap of the 10 Å illite and 7 Å chlorite peaks by the 10 Å muscovite and the 7 Å kaolinite peaks. Other factors associated, such as deformation, lithology, hydrothermal alteration and surface weathering may also affect Illite and chlorite (e.g., Hara and Kimura 2000). Hara and Kimura (2000) in particular, underlined that these factors can easily influence the $Hb_{rel}$ values near a fault zone, and thus $Hb_{rel}$ values and Ch peak-ratios produced here should be considered with great caution.

**[Please insert Table 2 here]**

*4.2.2. Arkitsa fault - Central Greece*

XRD bulk mineral compositions and clay mineral compositions of both the fault-rock samples and the host-rock sample from Arkitsa are summarized in Table 3. The bulk rock mineralogy showed a composition of calcite, with minor dolomite, quartz, and lesser amounts of clay minerals and goethite. In sample ARK3 (fault-rock sample) and ARKITSA (host-rock), dolomite could not be detected by XRD analysis. The presence of dolomite and quartz and the absence of talc in the two slickenside fault-rock samples suggest that alteration temperatures never exceeded about 400 ⁰C (e.g., Winkler, 1976 p. 114; Boulvais et al. 2006).

The dominant clay mineral phases (<2 μm) are illite/muscovite, chlorite/kaolinite and non-clay mineral quartz and goethite in all samples. Some minor saponite was also found in the host-rock sample. The presence of saponite, a trioctahedral mineral of the smectite group, may indicate a low temperature and pressure environment, as in general trioctahedral smectites do not persist to high temperatures (e.g., Eberl et al. 1978).

Similar to the fault-rock samples from Mattinata, based on SEM/EDX analysis and XRD peak shape at 7 Å, both kaolinite and chlorite most likely occur in all samples from Arkitsa, thus we here refer to both minerals. Further, since differentiation between illite and muscovite can only be achieved through XRD analysis on glycolated samples (not performed in this study), both illite/muscovite may be present in our samples (event though muscovite was not observed during SEM/EDX analysis). The presence of kaolinite and quartz indicates a maximum temperature of <300 ⁰C and most probably a low-temperature meteoric water-rock interaction. However, since pure smectite was not observed (based on SEM/EDX a mixed-layered illite-smectite is more probable), a temperature regime >225 ⁰C, causing smectite transformation to illite-smectite may be suggested.

**[Please insert Table 3 here]**

Hb$_{rel}$ values and Ch peak-ratios for all samples are shown in Table 4. The host-rock has an Hb$_{rel}$ value and Ch peak-ratio of 292 and 1.8 respectively, indicating a low diagenetic temperature. Hb$_{rel}$ values and Ch peak-ratios for the fault-rock samples varies from 167 to 298 for Hb$_{rel}$, and 0.9 to 2.0 for Ch peak-ratios, indicating metamorphism in anchizone or even epizone.

**[Please insert Table 4 here]**

*4.2.3. Kera fault - Western Crete*

The major mineral assemblages via bulk XRD analysis in all samples collected at Kera fault were identified as calcite, dolomite, quartz, phyllosilicate minerals, with minor illite/muscovite and chlorite/kaolinite (Table 5).

Similar to Mattinata and Arkitsa, XRD analysis on clay size separates (<2 μm) from the fault-rock samples from Kera revealed illite/muscovite, chlorite/kaolonite with some non-clay minerals (quatz and goethite) also present. Since Kera fault-rock samples show very similar mineralogy to Mattinata and Arkitsa samples, the same conclusion can be drawn regarding prevailing conditions and temperature regimes in Kera fault zone. In addition, like in Arkitsa fault, the presence of dolomite and quartz and the absence of talc in the fault-rock samples suggest that alteration temperatures never exceeded about 400 ºC.

Clay size separates (<2 μm) of the breccia sample consists of illite/muscovite and chlorite/kaolonite, as well as minor quartz and lesser cristobalite. The presence of cristobalite suggests a high temperature regime, as the conversion of quartz into the higher temperature forms of silica requires temperatures > 900 ºC (e.g., Ringdalen, 2014).

**[Please insert Table 5 here]**

Table 6 presents the Hb$_{rel}$ values and Ch peak-ratios for the various samples from Kera. The individual Hb$_{rel}$ values measured for the slickenside fault-rock samples range from 211 to 299 defining a temperature range which varies from the anchizone to the diagenetic field. The Ch peak-ratios for the same samples also indicate low-grade metamorphism in the anchizone as well as a diagenetic environment. Examination of the host-rock sample signified that it was heated below 250° C. Using $Hb_{rel}$, the breccia sample is placed in the diagenetic field, while the Ch peak-ratio indicated a palaeotemperature ranging from the achizone to epizone.

As mentioned above, the presence of cristobalite in the breccia sample indicates a temperature regime beyond 900 ºC, however such high temperatures are not expected for incohesive breccias, which their development is limited to the near-surface, between 1-4 km depth (Sibson, 1977). Therefore, we believe that the presence of cristobalite represents a secondary mineral, most probably precipitated into the breccia formation from circulating groundwater and not representing an authigenic mineral.

**[Please insert Table 6 here]**

## 5. Summary and conclusions

The fundamental premise of luminescence and ESR dating of past seismic events is that during the deformation of fault-rocks, their constituent minerals have been subject to various physical and chemical processes which allowed for the resetting of their luminescence/ESR signal. Therefore, understanding fault-rocks mineralogical and microstructural characteristics, together with their prevailing thermal regimes is crucial for establishing their suitability for luminescence/ESR dating. In the first phase of this research project, analyses of a number of samples collected from three carbonate fault zones yielded observations that support the suggestion that mirror-like cataclastic structures are good candidates for direct dating of past seismic activity.

Several microstructural observations on carbonate fault slickenside samples provided evidence of intense cataclastic deformation. Sections cut perpendicular to the slip surface of the slickensides revealed the presence of truncated clasts, providing evidence of intense strain localization during fast (seismic) slip. The slickenside fault-rock samples have experienced extensive fracturing which is demonstrated by the presence of calcite/quartz filled (early-stage) fractures and undeformed veins as well as broken apart (late co-seismic) and discontinuous veins. Further, the observation of quartz and calcite serrated grain boundaries indicated the effect of either pressure solution or of strain induced migration, both processes taken place under a tectonic manner. Additionally, the presence of stylolites at high angles on Kera fault-rock samples signified that pressure solution was also an important deformation mechanism in Kera fault zone.

Quartz grain overgrowths and the presence of conchoidal fractures on quartz grains are also common features in the carbonate fault-rock samples, indicating pressure on grains surface as well as powerful impact. Further, the occurrence of certain minerals associated with microstructural characteristics on quartz grains, indicated that fluid composition in the fault zones changed over time. High Si content is indicated when the quartz overgrowths developed while subsequent changes in fluids are indicated by the extensive corrosion of quartz grains and the development of etch pits on quartz overgrowths.

Examination of a fault-breccia sample collected from an outcrop adjacent to the Kera slickenside formation also revealed extensive deformation, in the form of highly crashed grains, antithetic microfractures in clasts as well as corroded quartz grains and quartz overgrowths. Furthermore, observations of the breccia sample showed that in places muscovite have been replaced by fine-grained aggregates of illite and kaolinite, indicating (late, after-brecciation) fluids circulation in a low temperature diagenetic environment.

Optical observations and EDX analysis of the calcareous host-rock samples revealed that they are typical limestones with a matrix consisting mainly of calcite grains, ooids and minor amounts of clay minerals. Host-rock samples from Mattinata fault show no signs of intense tectonic deformation, while the presence of veins in the host-rock samples from Arkitsa and Kera indicated some degree of tectonic deformation.

SEM-EDX investigation on fault-rock samples allowed the identification of minerals illite, chlorite, kaolinite and possibly illite-smectite mixed-layer phases. A range of illite crystal shapes can be observed by SEM in all slickenside fault-rock samples, however illite more commonly forms irregular flake-like platelets or subhedral to euhedral crystals. Chlorite appears in the form of platelets usually arranged perpendicular to one another or forming rosettes of individual platelet crystals. Further, kaolinite typically occurs as face-to-face stacks of pseudohexagonal plates or books.

XRD analysis of the bulk and clay size fractions of all fault-rock samples revealed that they have a similar mineralogy dominated by calcite, quartz and clay minerals as well as goethite. The low illite/muscovite content (bulk XRD) in all slickenside fault-rock samples may indicate a relatively low permeability and moderate intensity of fluid-rock interaction. Even though a low permeability is suggested, the occurrence of kaolinite may indicate hydrothermal episodes with more acidic conditions and leaching cations, or it may reflect late, minor alteration by low-temperature surface water.

Regarding palaeotemperature regimes, the presence of kaolinite and quartz in all slickenside fault-rock samples suggest that maximum temperature never exceeded approximately 300 ºC. Further, the presence of dolomite along with quartz and the absence of talc (Arkitsa and Kera fault-rock samples) suggest that alteration temperatures never exceeded 400 ºC.

The estimation of palaeotemperatures from $Hb_{rel}$ values for the fault-rock samples collected from Mattinata, Arkitsa and Kera also indicated a range which varies from the diagenetic field (<200 ºC) to the anchizone (200-370 ºC). In many cases palaeotemperatures derived from Ch peak-ratios were in fair agreement with palaeotemperatures using $Hb_{rel}$ values, however, there are instances where Ch peak-ratios indicated higher heating regimes (>400 ºC), contradicting $Hb_{rel}$ palaeotemperatures. We interpreted this discrepancy as a result of a possible overlap of the 7 Å chlorite peaks by the 7 Å kaolinite peaks. In the case of Arkitsa and Kera, host-rock

$Hb_{rel}$ values and Ch peak-ratios indicated that they have experienced temperatures of <250 ⁰C, thus adding to the robustness of the $Hb_{rel}$ and Ch peak-ratio approach, as host-rocks are expected to present no heating signatures. However, this is not the case of the host-rock samples collected in Mattinata, where both $Hb_{rel}$ values and Ch peak-ratios indicated a high temperature. We interpreted this inconsistency to be due to complications associated with possible XRD peaks overlaps.

This preliminary investigation has indicated that carbonate fault slickensides and breccia samples can be considered datable materials, since they contain suitable minerals (in our case quartz) for luminescence and ESR dating, have experienced repeated cataclastic deformation and have been subject to various physical and chemical processes as well as pressure and temperature conditions which probably allowed for the resetting of their luminescence/ESR signal. Having established the potential of our samples to be used for luminescence and ESR dating, the next step will be the examination of the complete resetting of their luminescence and ESR signals, which may include stress simulation experiments backed by luminescence and ESR measurements.

References

Ambraseys, N., Jackson, J., 1990. Seismicity and associated strain of central Greece between 1890 and 1988. Geophysical Journal International 101, 663-708.

Anthony, J.W., Bideaux, R.A., Bladh, K.W., Nichols, M.C., 2009. The Handbook of Mineralogy. Mineralogical Society of America, Chantilly, VA, USA.

Árkai, P., 1991. Chlorite crystallinity: An empirical approach and correlation with illite crystallinity, coal Tank and mineral facies as exemplified by Palaeozoic and Mesozoic rocks of northeast Hungary. Journal of Metamorphic Geology 9, 723-734.

Banerjee, D., 1996. New Applications of Thermoluminescence. Ph.D. thesis, Gujarat University, India.

Banerjee, D., Singhvi, A.K., Pande, K., Gogte, V.D., Chandra, B. P., 1999. Towards a direct dating of fault gouges using luminescence dating techniques - methodological aspects. Current Science 77, 256-269.

Barthelmy, D., 2010. The mineralogy database. Available at: http://webmineral.com/. Accessed: 11/2017.

Bense, F.A., Wemmer, K., Löbens, S., Siegesmund, S., 2014. Fault gouge analyses: K-Ar illite dating, clay mineralogy and tectonic significance-a study from the Sierras Pampeanas, Argentina. International Journal Earth Sciences 103, 189-218.

Biscaye, B.E., 1964. Distinction between kaolinite and chlorite in recent sediments by X-ray diffraction. The American Mineralogist 49, 1281-1289.

Bjorevkke, K., Malm, O., Elveruoi A., 1979. Diagenesis in Mesozoic sandstones from Spitsbergen and the North Sea-a comparison. Geologische Rundschau 68, 1151-1171.

Blanco, A., Dee Tomasi, F., Filippo, E., Manno, D., Perrone, M.R., Serra, A., Tafuro, A.M., Tepore, A., 2003. Characterization of African dust over southern Italy. Atmospheric Chemistry and Physics 3, 2147-2159.

Blum, A.E., Yund, R.A., Lasaga, A.C., 1990. The Effect of dislocation density on the dissolution rate of quartz. Geochimica et Cosmochimica Acta 54, 283-297.

Boneh, Y., Sagy, A., Reches, Z., 2013. Frictional strength and wear-rate of carbonate faults during high-velocity, steady-state sliding. Earth and Planetary Science Letters 381, 127-137.

Boulvais, P., de Parseval, P., D'Hulst, A., Paris, P., 2006. Carbonate alteration associated with talc-chlorite mineralization in the eastern Pyrenees, with emphasis on the St. Barthelemy Massif. Mineralogy and Petrology 88, 499-526.

Brankman, C., and Aydin, A. 2004. Uplift and contractional deformation along a segmented strike-slip fault system: The Gargano Promontory, southern Italy. Journal of Structural Geology 26, 807-824.

Brankman, C.M., Aydin, A., 2004. Uplift and contractional deformation along a segmented strike-slip fault system: the Gargano Promontory, southern Italy. Journal of Structural Geology 26, 807-824.

Brantley, S.L., Crane, S.R., Crerar, D.A., Hellman, R., Stallard, R., 1986. Dissolution at dislocation etch pits in quartz. Geochimica et Cosmochimica Acta 50, 2349-2361.

Burley, S.D., 1984. Distribution and origin of authigenic minerals in the Triassic Sherwood Sandstone Group, UK. Clay Minerals 19, 403-441.

Caine, J.S., Evans, J.P., Forster, C.B., 1996. Fault zone architecture and permeability structure. Geology 24, 1025-1028.

Caputo, S., Catalano, C., Monaco, G., Romagnoli, G., Tortorici, L., 2010. Active faulting on the island of Crete (Greece). Geophysical Journal International 183, 111-126.

Chen, X., Madden, A.S., Bickmore, B.R., Reches, Z., 2013. Dynamic weakening by nanoscale smoothing during high-velocity fault slip. Geology 41, 739-742.

Chester, F.M., Logan, J.M., 1986. Implications for mechanical properties of brittle faults from observations of the Punchbowl Fault Zone, California. Pure and Applied Geophysics 124, 79-106.

Dalla Torre, M., de Capitani, C., Frey, M., Underwood, M.B., Mullis, J., Cox, C., 1996. Very-low temperature metamorphism of shales from the Diablo Range, Franciscan complex, California: new constraints on the exhumation path. Geological Society of America Bulletin 108, 578-601.

De Paola, N., 2013. Nano-powder coating can make fault surfaces smooth and shiny: implications for fault mechanics? Geology 41, 719-720.

Desbois, G., Höhne, N., Urai, J.L., Bésuelle, P., Viggiani, G., 2017. Deformation in cemented mudrock (Callovo-Oxfordian Clay) by microcracking, granular flow and phyllosilicate plasticity: insights from triaxial deformation, broad ion beam polishing and scanning electron microscopy. Solid Earth 8, 291-305.

Di Toro, G., Goldsby, D.L., Tullis, T.E., 2004. Friction falls towards zero in quartz rock as slip velocity approaches seismic rates. Nature 427, 436-439.

Doglioni, C., Mongelli, F., Pieri, P., 1994, The Puglia uplift (SE Italy): An anomaly in the foreland of the Apenninic subduction due to buckling of a thick continental lithosphere. Tectonics 13, 1309-1321.

Dunoyer de Segonzac, G., 1970. The transformation of clay minerals during diagenesis and low-grade metamorphism: a review. Sedimentology 15, 281-346.

Dunoyer de Segonzac, G., Ferrero, J., Kubler, B., 1968. Sur la cristallinite de L'illite dans la diagenese et l'anchimetamorphisme. Sedimentology 10, 137-143.

Eberl, D.D., Whitney, G., Khoury, H., 1978. Hydrothermal reactivity of smectite. American Mineralogist 63, 401-409.

Evans, J.P, Chester, F.M., 1995. Fluid-rock interaction in faults of the San Andreas System: inferences from San Gabriel fault rock geochemistry and microstructures. Journal of Geophysical Research 100, 13007-13020.

Fantong, E.B., Takeuchi, A., Doke, R., 2013. Electron spin resonance (ESR) dating of calcareous fault gouge of the Ushikubi fault, central Japan. Applied Magnetic Resonance 44, 1105-1123.

Favali, P., Funiciello, R., Mattietti, G., Mele, G., Salvini, F., 1993. An active margin across the Adriatic Sea (central Mediterranean Sea). Tectonophysics 219, 109-117.

Fondriest, M., Smith, S.A.F., Candela, T., Nielsen, S., Mair, K., Di Toro, G., 2013. Mirror-like faults and power dissipation during earthquakes. Geology 41, 1175-1178.

Fossen, H., 2010. Structural geology. Cambridge University Press, Cambridge, UK.

Fukuchi, T., Imai, N., 1998. Resetting experiment of E´ centers by natural faulting-the case of the Nojima earthquake fault in Japan. Quaternary Science Reviews 17, 1063-1068.

Funiciello, R., Montone, P., Salvini, F., Tozzi, M., 1988. Caratteri strutturali del Promontorio del Gargano. Memorie - Società Geologica Italiana 41, 1235-1243.

Ganas, A., Roberts, G., Memou, P., 1998. Segment boundaries, the 1894 ruptures and strain patterns along the Atalanti fault, central Greece. Journal of Geodynamics 26, 461-486.

Hadizadeh, J., 1994. Interaction of cataclasis and pressure solution in a lowtemperature carbonate shear zone. Pure and Applied Geophysics 143, 255-280.

Hara, H., Kimura, K., 2000. Estimation of errors in measurement of illite crystallinity: the limits and problems of application to accretionary complexes. The Journal of the Geological Society of Japan 106, 264-279. (in Japanese with English abstract).

Hashimoto, Y., Ujiie, K., Sakaguchi, A., Tanaka, H., 2007. Characteristics and implication of clay minerals in the northern and southern parts of the Chelungpu fault, Taiwan. Tectonophysics 443, 233-242.

Hobbs, B.H., Means W.D., Williams, P.F., 1976. An outline of structural geology. John Wiley and Sons, New York, USA.

Isaacs, A.J., Evans, J.P., Song, S.R., Kolesar, P.T., 2007. Structural, mineralogical, and geochemical characterization of the Chelungpu Thrust Fault, Taiwan. Terrestrial Atmospheric and Oceanic Sciences 18, 183-221.

Jackson, J., McKenzie, D., 1999. A hectare of fresh striations on the Arkitsa fault, central Greece. Journal of Structural Geology 21, 1-6.

Jessell, M.W., 1986. Grain boundary migration and fabric development in experimentally deformed octachloropropane. Journal of Structural Geology 8, 527-542.

Johnson, P.A., Hochella, M.F., Parks, J.A., Blum, A.E., 1992. Direct observation of muscovite basal plane dissolution and secondary phase formation. An XPS, LEED, and SFM study. In: Kharaka, Y. K., Maest, A. S. (Eds.), Water-rock interaction. Balkema, Rotterdam. 159-162.

Johnson, N.M., Stix, J., Tauxe, L., Cerveny, P.F., Tahirkheli, R.A.K., 1985. Paleomagnetic chronology, fluvial processes and tectonic implications of the Siwalik deposits near Chinji Village, Pakistan. Journal of Geology 93, 27-40.

Jolivet, L., Goffe, B., Monie, P., Truffert-Luxey, C., Bonneau, M., 1996. Miocene detachment in Crete and exhumation PT- t paths of high-pressure metamorphic rocks. Tectonics 15, 1129-1153.

Jones, R.R., Kokkalas, S., McCaffrey, K.J.W., 2009. Quantitative Analysis and Visualisation of non-planar fault surfaces using Terrestrial Laser Scanning ("LiDAR")-the Arkitsa Fault, central Greece as a case study. Geosphere 5, 465-482.

Joshi, M.S., Kotru, P.N., 1969. Hydrothermal etching of matched fractured prism faces of natural quartz. Soviet Physics: Crystallography 14, 427-428.

Kantorowicz, J., 1984. Nature, origin and distribution of authigenic clay minerals from Middle Jurassic clastic sediments, Ravenscar and Brent Group sandstones. Clay Minerals 19, 359-377.

Kilias, A., Fassoulas, C., Mountrakis, D., 1994. Tertiary extension of continental crust and uplift of Psiloritis metamorphic core complex in the central part of the Hellenic Arc (Crete, Greece). Geologische Rundschau 83, 417-430.

Kitagawa, Y., Fujimori, K., Koizumi, N., 2007. Temporal changes in permeability of the Nojami fault zone by repeated water injection experiments. Tectonophysics 443, 183-192.

Kokkalas, S., Jones, R.R., McCaffrey, K.J.W., Clegg, P., 2007. Quantitative fault analysis at Arkitsa, Central Greece, using terrestrial laser-scanning (LiDAR). Bulletin of the Geological Society of Greece 40, 1959-1972.

Kragelskii, I.V., 1965. Friction and Wear. Butterworths, London.

Kübler, B., 1968. Evaluation quantitative du métamorphism par la cristallinité de l'illite. Bulletin du Centrem de Recherche de Pau-SNPA, 2: 385-397.

Kübler, B., 1967. La cristallinitede l'illite et les zones tout a fait superieures de metamorphisme. In: Schaer, J. P., (Eds.), Colloque sur les etages tectonique. Éditions de la Baçonnière, Neuchatel, Switzerland. 105-122.

Le Ribault, L., 1977. L'exoscopie des quartz. Editions Masson, Paris.

Lin, A., 2008. Fossil Earthquakes: The Formation and Preservation of Pseudotachylytes. Springer-Verlag, Berlin.

Makropoulos, K.C., Kouskouna, V., 1994. The Ionian Islands earthquakes of 1767 and 1769: seismological aspects. Contribution of historical information to a realistic seismicity and hazard assessment of an area, in Historical Investigation of European Earthquakes, Materials of the CEC Project: Review of historical Seismicity in Europe. National Research Council-Milano 2, 27-36.

Margolis, S.V., Krinsley, D.H., 1974. Processes of formation and environmental occurrence of microfeatures of detrital quartz grains. American Journal of Science 274, 449-464.

Martinis, B., Pieri, M., 1964. Alcune notizie sulla formazione evaporitica dell'Italia centrale e meridionale. Memorie - Società Geologica Italiana 4, 649-678.

Merriman, R.J., Roberts, B., Peacor, D.R., 1990. A transmission electron microscope study of white mica crystallite size distribution in a mudstone to slate transitional sequence, North Wales, UK. Contributions to Mineralogy and Petrology 106, 27-40.

Milliken, K.L., 2003. Microscale distribution of kaolinite in Breathitt Formation sandstones (middle Pennsylvanian): implications for mass balance. In: Worden, R., Morad, S. (Eds.), Clay mineral cements in sandstones. International Association of Sedimentologists Special Publication 34, 343-360.

Moore, D.M., Reynolds, R.C., 1997. X-Ray Diffraction and the Identification and Analysis of Clay Minerals. Oxford University Press, Oxford.

Morrow, C.A, Moore, D.E, Lockner, D.A., 2001. Permeability reduction in granite under hydrothermal conditions. Journal of Geophysical Research 106, 30551-30560.

Mountrakis, D., Kilias, A., Pavlaki, A., Fassoulas, Ch., Thomaidou, E., Papazachos, C., Papaioannou, Ch., Roumelioti, Z., 2012. Neotectonic study of Western Crete and implications for seismic hazard assessment. Journal of the Virtual Explorer 42, doi:10.3809/jvirtex.2011.00285.

Nishimura, S., Horinouchi, T., 1989. Thermoluminescence ages of some quartz in fault gouges. Journal of Physics of the Earth 37, 313-323.

Offler, R., Och, D.J., Phelan, D., Zwingmann, H., 2009. Mineralogy of gouge in north-northeast-striking faults, Sydney region, New South Wales. Australian Journal of Earth Sciences 56, 889-905.

Ortolani, F., Pagliuca, S., 1988. Il Gargano (Italia meridionale): un settore di 'avampaese' deformato tra la catena appenninica e dinarica. Memorie - Società Geologica Italiana 41, 1245-1252.

Ostrom, M.E., 1961. Separation of clay minerals from carbonate rocks by using acids. Journal of Sedimentary Petrology 31, 123-129.

Papazachos, B., Papazachou, C., 1997. The Earthquakes of Greece. Ziti Publications, Thessaloniki.

Papoulis, D., Romiou, D., Kokkalas, S., Lampropoulou, P., 2013. Clay minerals from the Arkitsa fault gouge zone, in Central Greece, and implications for fluid flow. Bulletin of the Geological Society of Greece 48, 616-624.

Piccardi, L., 2005. Paleoseismic evidence of legendary earthquakes: The apparition of Archangel Michael at Monte Sant'Angelo (Italy). Tectonophysics 408, 113-128.

Pe-Piper, G., Panagos, A.G., 1989. Geochemical characteristics of the triassic volcanic rocks of Evia: petrogenetic and tectonic implications. Ofioliti 14, 33-50.

Poirier, J.P., Guillope, M., 1979. Deformation induced recrystallization of minerals. Bulletin de Minéralogie 102, 67-74.

Ratcliffe, N.M., Burton, W.C., 1988. Structural analysis of the Furlong fault and the relationship of mineralization to faulting and diabase intrusion, Newark basin, Pennsylvania. In: Froelich, A. J., Robinson, G. R. (Eds.), Studies of the Early Mesozoic Basins of the Eastern United States. U.S. Geological Survey Bulletin, 176-193.

Rhodes, E.J., Singarayer, J.S., Raynal, J.P., Westaway, K.E., Sbihi-Alaoui, F.Z., 2006. New age estimates for the Palaeolithic assemblages and Pleistocene succession of Casablanca, Morocco. Quaternary Science Reviews 25, 2569-2585.

Rimstidt, J.D., Barnes, H.L., 1980. The kinetics of silica-water reactions. Geochimica et Cosmochimica Acta 44, 1683-1699.

Ringdale, E., 2014. Changes in quartz during heating and the possible effects on si production. The Journal of The Minerals, Metals and Materials Society (JOM) 67, 484-492.

Salvini, F., Billi, A., Wise, D.U., 1999. Strike-slip fault-propagation cleavage in carbonate rocks: the Mattinata Fault zone, Southern Apennines, Italy. Journal of Structural Geology 21, 1731-1749.

Schleicher, A.M., Warr, L.N., van der Pluijm, B.A., 2009. On the origin of mixed-layered clay minerals from the San Andreas Fault at 2.5-3 km vertical depth (SAFOD drillhole at Parkfield, California). Contribution to Mineralogy and Petrology 157, 173-187.

Sawakuchi, A.O., Mendes, V.R, Pupim, F.N., Mineli, T.D, Ribeiro, L.M.A.L., Zular, A., Guedes, C.C.F., Giannini, P.C.F., Nogueira, L, Sallun Filho, W., Assine, M.L., 2016. Optically stimulated luminescence and isothermal thermoluminescence dating of high sensitivity and well bleached quartz from Brazilian sediments: from Late Holocene to beyond the Quaternary? Brazilian Journal of Geology 46, (Suppl. 1), 209-226.

Scholz, C.H., 2002. The mechanics of earthquakes and faulting. Cambridge University Press, Cambridge, UK. 471.

Shen, J., Yang, W., Liu, T., Huang, X., Zheng, W., Yu, L., Wang, G., 2014. Dating fault activity based on surface texture of quartz grains from the Bailong river fault. Acta Geologica Sinica 88, 1131-1144.

Sibson, R.H., 2003. Thickness of the seismic slip zone. Bulletin of the Seismological Society of America 93, 1169-1178.

Sibson, R.H., 1977. Fault rocks and fault mechanisms. Journal of the Geological Society 133, 191-213.

Siman-Tov, S., Aharonov, E., Boneh, Y., Reches, Z., 2015. Fault mirrors along carbonate faults: formation and destruction during shear experiments. Earth and Planetary Science Letters 430, 367-376.

Siman-Tov, S., Aharonov, E., Sagy, A., Emmanuel S., 2013. Nanograins form carbonate fault mirrors. Geology 41, 703-706.

Singhvi, A.K., Banerjee, D., Pande, K., Gogte, V., Vadiya, K.S., 1994. Luminescence studies on neotectonic events in South-Central Kumaun Himalaya-a feasibility study. Quaternary Science Reviews 13, 595-600.

Smith, S.A.F., Billi, A., Di Toro, G., Spiess, R., 2011. Principal Slip Zones in Limestone: Microstructural Characterization and Implications for the Seismic Cycle (Tre Monti Fault, Central Apennines, Italy). Pure and Applied Geophysics 168, 2365-2393.

Sone, H., Yeh, E., Nakaya, T., Hung, J., Ma, K., Wang, C., Song, S., Shimamoto, T., 2007. Mesoscopic structural observations of cores from the Chelungpu fault system, Taiwan Chelungpu Fault Drilling Project, Hole-A, Taiwan. Terrestrial Atmospheric and Oceanic Sciences 18, 2-19.

Spry, A., 1969. Metamorphic Textures. Pergamon Press Ltd., Oxford.

Stiros, S., 2001. The AD 365 Crete earthquake and possible seismic clustering during the fourth to sixth centuries AD in the Eastern Mediterranean: a review of historical and archaeological data. Journal of Structural Geology 23, 545-562.

Surma, F., Geraud, Y., 2003. Porosity and Thermal Conductivity of the Soultz-sous-Forets Granite. Pure and Applied Geophysics 160, 1125-1136.

Ten Veen, J.H., Kleinspehn, K.L., 2003. Incipient continental collision and plate-boundary curvature: Late Pliocene-Holocene transtensional Hellenic fore arc, Crete, Greece. Journal of the Geological Society 160, 1-6.

Tondi, E., Piccardi, L., Cacon, S., Kontny. B., Cello, G., 2005. Structural and time constraints for dextral shear along the seismogenic Mattinata Fault (Gargano. southern Italy). Journal of Geodynamics 40, 134-152.

Toyoda, S., Rink, W.J., Schwarcz, H.P., Rees-Jones, J., 2000. Crushing effects on TL and OSL on quartz: relative to fault dating. Radiation Measurements 32, 667-672.

Tsakalos, E., 2016. Geochronology and exoscopy of quartz grains in environmental determination of coastal aeolianites in SE Cyprus. Journal of Archaeological Science: Reports 7, 679-686.

Tsakalos, E., Athanassas, C., Bassiakos, Y., 2015. Luminescence dating and quartz grain surface features of aeolian sediments of a site in southeast Cyprus. 6th Symposium of the Hellenic Society for Archaeometry. British Archaeological Reports, 201-206.

Tsakalos, E., Athanassas, C., Bassiakos, Y., Tsipas, P., Triantaphyllou, M., Geraga, M., Papatheodorou, G., Christodoulakis, J., Kazantzaki, M., 2016. Luminescence geochronology and paleoenvironmental implications of coastal deposits of southeast Cyprus. Journal of Archaeological and Anthropological Sciences 10, 41-60.

Urai, J.L., Means, W.B., Lister, G.S., 1986. Dynamic recrystallization of mineral. Geophysical Monograph-American Geophysical Union 36, 161-199.

Verberne, B.A., Plümper, O., de Winter, D.A.M., Spiers, C.J., 2014. Rock mechanics. Superplastic nanofibrous slip zones control seismogenic fault friction. Science 346, 1342-1344.

Voll, G., 1976. Recrystallization of Quartz, Biotite, Feldspars from Erstfeld to the Leventina Nappe, Swiss Alps, and its Geological Significance. Schweizerische Mineralogische und Petrographische Mitteilungen 56, 641-647.

Vos, K., Vandenberghe, N., Elsen, J., 2014. Surface textural analysis of quartz grains by scanning electron microscopy (SEM): From sample preparation to environmental interpretation. Earth-Science Reviews 128, 93-104.

Vrolijk, P., van der Pluijm, B.A., 1999. Clay gouge. Journal of Structural Geology 21, 1039-1048.

Wallace, R.E., Morris, H.T., 1986. Characteristics of faults and shear zones in deep mines. Pure and Applied Geophysics 124, 107-125.

Wang, H., Frey, M., Stern, W.B., 1996. Diagenesis and metamorphism of clay minerals in the Helvetic Alps of Eastern Switzerland. Clays and Clay Minerals 44, 96-112.

Weaver, C.E., 1960. Possible uses of clay minerals in search for oil. Bulletin - American Association of Petroleum Geologists 44, 1505-1518.

Weber, K., 1972. Notes on the determination of illite crystallinity. Neues Jahrbucl Mineralisches Monatshefte 6, 267-276.

White, S.E., 1977. Geological significance of recovery and recrystallization processes in quartz. Tectonophysics 39, 143-170.

Wilson, B., Dewers, T., Reches, Z., Brune, J., 2005. Particle size and energetics of gouge from earthquake rupture zones. Nature 434, 749-752.

Winkler, H.G.F., 1976. Petrogenesis of metamorphic rocks. Springer-Verlag, New York.

Wu, F.T., Blatter, L., Roberson, H., 1975. Clay gouges in the San Andreas fault system and their possible implications. Pure and Applied Geophysics 113, 87-95.

Yang, H.L., Chen, J., Yao, L., Liu, C.R., Shimamoto, T., Thompson Jobe, J.A., 2018. Resetting of OSL/TL/ESR signals by frictional heating in experimentally sheared quartz gouge at seismic slip rates. Quaternary Geochronology, https://doi.org/10.1016/j.quageo.2018.05.005.

Yates, D.M., Rosenberg, P.E., 1997. Formation and stability of end-member illite: II. Solid equilibration experiments at 100 to 250 ºC and Pv,solution. Geochimica et Cosmochimica Acta 61, 3135-3144.

Figure Captions

Fig. 1 Simplified geological sketch map of the Gargano Promontory, Italy with main fault patterns showing the location of the study area (red frame). Note the variability of depositional environments in the Promotory (from platform in west to basin in east). The inserted picture shows a fault slickenside of the main Mattinata fault on the NE slope of the Monte Saraceno.

Fig. 2 Geological setting of the Arkitsa-Atalanti area, central Greece, showing the location of the study area (red frame). The inserted photograph shows the fault slickenside of Arkitsa

Fig. 3 Geological map showing the major segments of the active fault zones and the study area (red frame). Kera fault is the shorter segment in the middle of the map, with a NE-SW orientation. The inserted picture shows the study site; a fault slickenside of the Kera fault.

Fig. 4 Representative photomicrographs of the slickenside fault-rock samples from Mattinata. a) A fault-rock sample cut parallel to slickenside and perpendicular to the slip zone. The slip surface is covered with quaternary material. The slip zone contains cataclasites, beneath a distinct slip layer which separates the slip zone from the slip surface. b) Sub-layers 1 and 2 within the slip surface. Note the occurrence of an uninterrupted corroded calcite vein that makes the boundary between the two sub-layers and suggesting fluid involvement. c) An area

in the slip zone with a relatively thick calcite vein with veinlets. d) A calcite vein showing blocky morphology and twinning.

Fig. 5 Observations on fault-rock and host-rock samples from Mattinata. a) Clay minerals illite and kaolinite forming grainsize patches and aggregates. b) Quartz exhibiting overgrowths (white arrows), dissolution (etch) pits (black arrows). c) Corrosion of a quartz grain associated with kaolinite/illite. d) Photomicrograph of the host-rock (limestone) showing a matrix of micritic calcite and skeletal fragments. e) An elongated quartz grain in the host-rock, surrounded by a matrix of calcite. I is illite, K kaolinite, Q quartz.

Fig. 6 SEM imagery and associated EDX spectrum of clay samples found in Mattinata fault-rocks. a) Illite making asymmetrical flake-like platelets sitting parallel to each other. b) Probable agglomerates of smectite-rich interlayered I-S. c) Chlorite forming rosettes of individual platelet crystals. d) Kaolinite in the form of face-to-face stacks of pseudohexagonal plates.

Fig. 7 Representative thin section microstructure of fault zone rock samples collected from the slickenside formation of Arkitsa. a) A fault-rock sample cut parallel to slickenside and perpendicular to the slip zone containing angular to sub-rounded clasts in a fine-grained, iron oxide impregnated, matrix, beneath a distinct slip layer which separates the slip zone from the slip surface. b) The ultracataclastic slip surface showing high iron oxide impregnation. c) Dolomite rhombs (white arrows) in the slip surface, surrounded by a fine grained calcite matrix. d) Extensive network of calcite-quartz veins in the slip zone. e) A vein in the slip zone, appearing heavily twinned and sheared with irregular, serrated grain boundaries. f) Recrystallized quartz grains (white arrows) at the edges of deformed quartz grains.

Fig. 8 Representative thin section photomicrographs on the host-rock limestone sample collected in Arkitsa. a) Calcite-micritic matrix rich in ooids. b) A vein network filled with sparry calcite grains.

Fig. 9 Representative SEM images of a fault-rock sample from Arkitsa. a) Clay minerals and dolomite in the form of patches and minor quartz grains. b) A highly corroded quartz grain. D is dolomite, K kaolinite, Q quartz, I illite.

Fig. 10 SEM images showing clay minerals in the fault-rock samples of Arkitsa, backed by EDX analysis. a) Illite subhedral to euhedral crystals with characteristic irregular edges. b) A possible illite-smectite mixed-layer with highly variable morphology. c) Chlorite platelets arranged in an almost a rosette pattern. d) Kaolinite forming face-to-face stacks of pseudohexagonal plates.

Fig. 11 Characteristic thin-section photomicrographs of fault-rock samples from Kera slickenside. a) A fault-rock sample cut parallel to slickenside and perpendicular to the slip zone showing the slip surface, slip layer and slip zone. b) Stylolite dissolution features in the slip zone. c) A calcite vein (white arrows) in the slip zone, cut by open fractures. d) Calcite within a vein showing a heterogeneous grain size distribution, undulose extinction and moderate twinning. e) Variably sized fragments of mainly quartz, dolomite and calcite aggregates, in a calcite matrix showing partial iron oxide impregnation. f) A calcite grain in the slip surface with highly variable, irregular and partially serrated boundaries.

Fig. 12 Photomicrographs and backscattered electron microscopy images of the host-rock and fault-breccia sample from Kera. a) Host-rock (limestone) composed almost exclusively by indeterminable abraded/rounded carbonate grains and ooids. A highly corroded calcite vein can also be observed (white arrows). b) Dolomite lenses with sutured boundaries in host-rock. c) Medium-crushed fault-breccia, composed of calcite, quartz, opaque minerals and muscovite. b) Antithetic microstructure of a calcite clast in fault-breccia. e) Replacement of muscovite by illite (white arrows) and kaolinite (black arrows) in fault-breccia sample.

Fig. 13 SEM images showing, a) a quartz grain from a slickenside fault-rock sample and b) breccia sample exhibiting overgrowths (black arrows) and dissolution (etch) pits (white

arrows) of varying density, size and shape. c) Conchoidal fractures (white arrows) and overgrowths (black arrows) on a quartz grain from a slickenside fault-rock sample and (d) on breccia sample.

Fig.14 Backscattered electron microscopy and EDX analysis on the fault-rock samples of Arkitsa showing clay minerals. a) Illite appearing as filamentous aggregates. b) Smectite or illite-smectite having a thin webby mineral morphology. c) Chlorite in the form of dispersed platelets arranged perpendicular to one another. d) Kaolinite developing as platelets.

Tables

Table 1 Mineralogical composition of fault-rock and host-rock samples from Mattinata.

|  |  | Bulk rock | | | | | Clay fraction | | | | |
|---|---|---|---|---|---|---|---|---|---|---|---|
|  |  | Illite/ Muscovite | Chlorite/ Kaolinite | Goethite | Quartz | Calcite | Illite/ Muscovite | Chlorite/ Kaolinite | Goethite | Quartz | Microcline |
| MAT2 | Fault-rock | * |  | * | ** | *** | ** |  | * | ** |  |
| MAT3 | Host-rock |  | * |  | * | *** | * | * |  | * |  |
| MAT5 | Fault-rock | * | * |  | * | *** | * | * | * | ** |  |
| MAT7 | Fault-rock | * | * | * | * | *** | * | * | * | * | * |
| MAT8 | Host-rock | * | * | * | * | *** | * | * | * | ** |  |

*** very abundant, ** abundant, * present

**Table 2** Illite $Hb_{rel}$ and Ch peak-ratios for the fault-rock and host-rock samples from Mattinata.

| Sample | Type | $Hb_{rel}$ | Inferred thermal excursion °C | Ch peak -ratio | Inferred thermal excursion °C |
|---|---|---|---|---|---|
| MAT2 | Fault-rock | 162 | 200-370 | - | - |
| MAT3 | Host-rock | 220 | 200-370 | 1.0 | >400 |
| MAT5 | Fault-rock | 231 | 200-370 | 0.9 | >400 |
| MAT7 | Fault-rock | 238 | 200-370 | 1.9 | 250-400 |
| MAT8 | Host-rock | 110 | >370 | 2.0 | 250-400 |

Table 3 Mineralogical composition of fault-rock and host-rock samples from Arkitsa.

|  |  | Bulk rock | | | | | | Clay fraction | | | | |
|---|---|---|---|---|---|---|---|---|---|---|---|---|
|  |  | Illite/ Muscovite | Chlorite/ Kaolinite | Goethite | Quartz | Calcite | Dolomite | Illite/ Muscovite | Chlorite/ Kaolinite | Goethite | Quartz | Saponite |
| ARK | Fault-rock | * | * | * | * | *** | * | ** | ** | * | ** |  |
| ARK1 | Fault-rock | * | * | ** | * | *** | ** | ** | ** | * | ** |  |
| ARK3 | Fault-rock | * | * | * | * | *** |  | * | * | * | ** |  |
| ARKITSA | Host-rock | * | * | * | * | *** |  | ** | * | * | ** | ** |

*** very abundant, ** abundant, * present

Table 4 Illite Hbrel and Ch peak-ratios for the fault-rock and host-rock samples from Arkitsa.

| Sample | Type | $Hb_{rel}$ | Inferred thermal excursion °C | Ch peak-ratio | Inferred thermal excursion °C |
|---|---|---|---|---|---|
| **ARKITSA** | Host-rock | 292 | <200 | 1.8 | <250 |
| **ARK** | Fault-rock | 167 | 200-370 | 2.0 | 250-400 |
| **ARK1** | Fault-rock | 181 | 200-370 | 0.9 | >400 |
| **ARK3** | Fault-rock | 298 | <200 | 1.9 | 250-400 |

Table 5 Mineralogical composition of fault-rock, breccia and host-rock samples from Kera.

| | | Bulk rock | | | | | Clay fraction | | | | |
|---|---|---|---|---|---|---|---|---|---|---|---|
| | | Illite/Muscovite | Chlorite/Kaolinite | Goethite | Quartz | Calcite | Dolomite | Illite/Muscovite | Chlorite/Kaolinite | Goethite | Quartz | Cristobalite |
| **KR1** | **Fault-rock** | * | * | | * | *** | * | * | * | ** | *** | |
| **KR2** | **Fault-rock** | * | | * | * | *** | ** | ** | * | * | ** | |
| **KR4A** | **Fault-rock** | * | * | | ** | *** | * | ** | ** | * | *** | |
| **KR7** | **Breccia** | ** | ** | * | *** | ** | | ** | ** | * | *** | * |
| **KR8** | **Host-rock** | * | | * | ** | *** | * | ** | * | ** | ** | |

\*** very abundant, ** abundant, * present

Table 6 Illite $Hb_{rel}$ and Ch peak-ratios for the fault-rock, breccia and host-rock samples from Kera.

| Sample | Type | $Hb_{rel}$ | Inferred thermal excursion °C | Ch. peak-ratio | Inferred thermal excursion °C |
|---|---|---|---|---|---|
| **KR1** | Fault-rock | 211 | 200-370 | 1.5 | 0-250 |
| **KR2** | Fault-rock | 299 | <200 | 1.4 | 0-250 |
| **KR4A** | Fault-rock | 297 | <200 | 1.7 | 0-250 |
| **KR7** | Breccia | 291 | <200 | 2.1 | 250-400 |
| **KR8** | Host-rock | 304 | <200 | 1.8 | 0-250 |

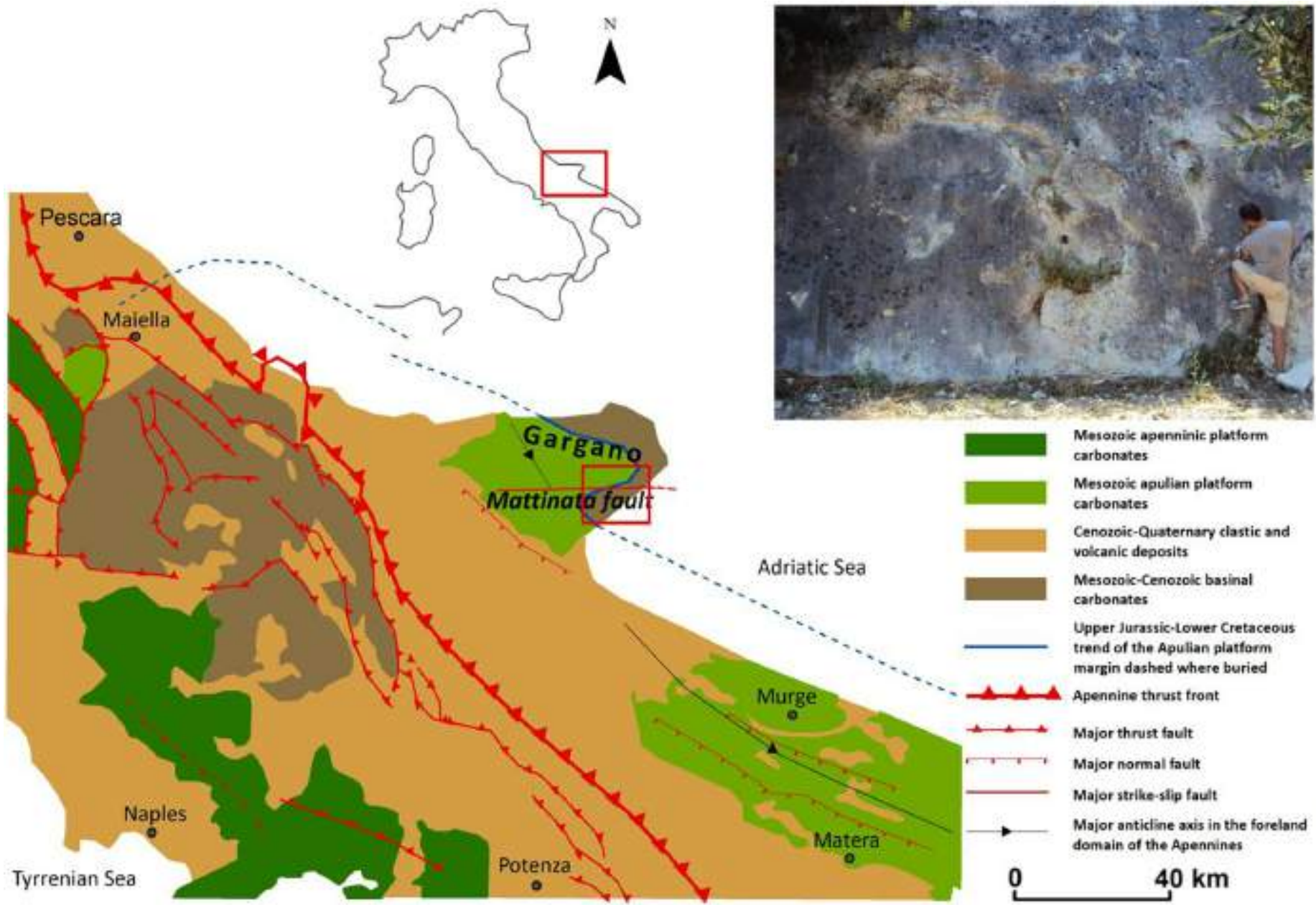

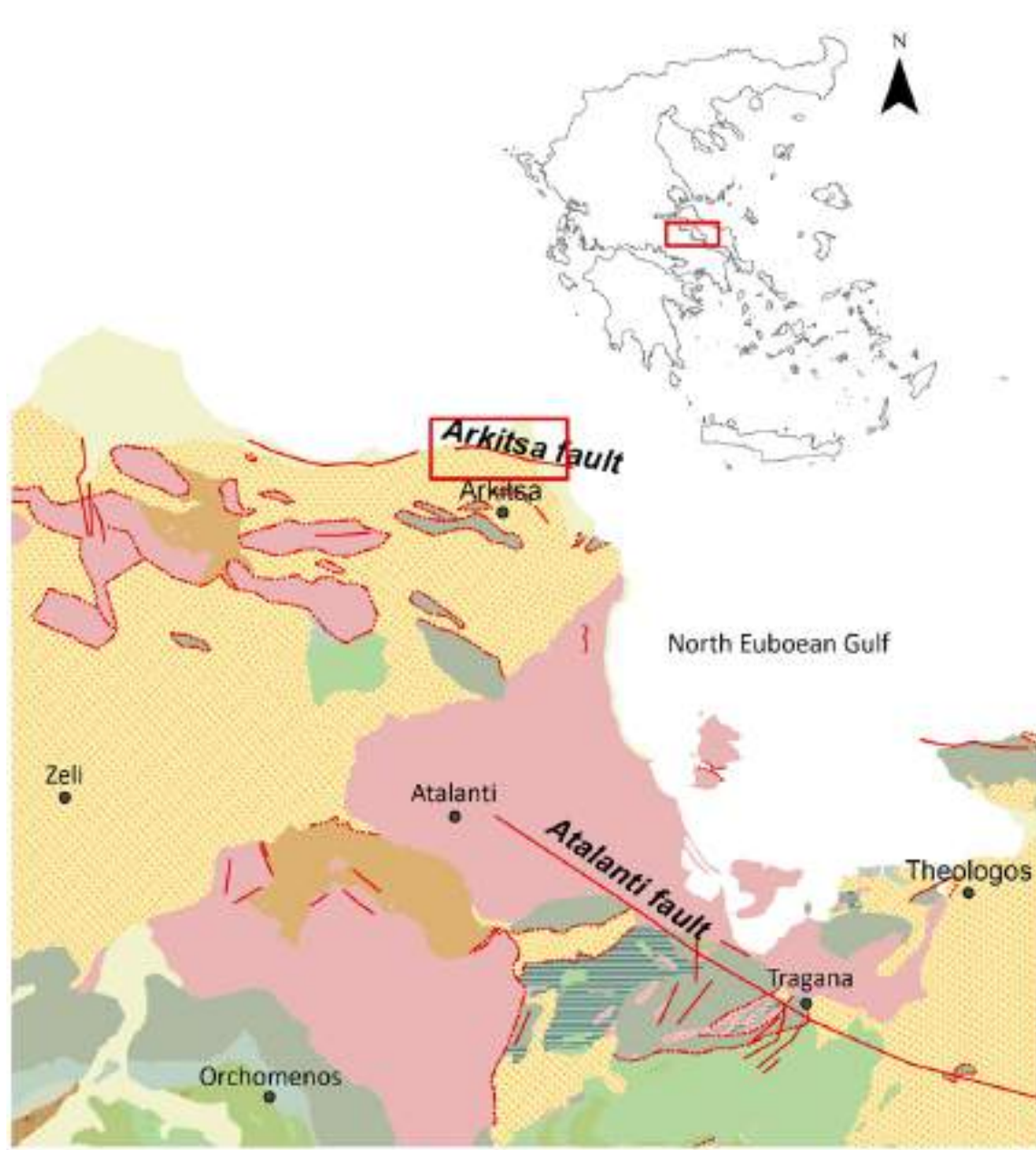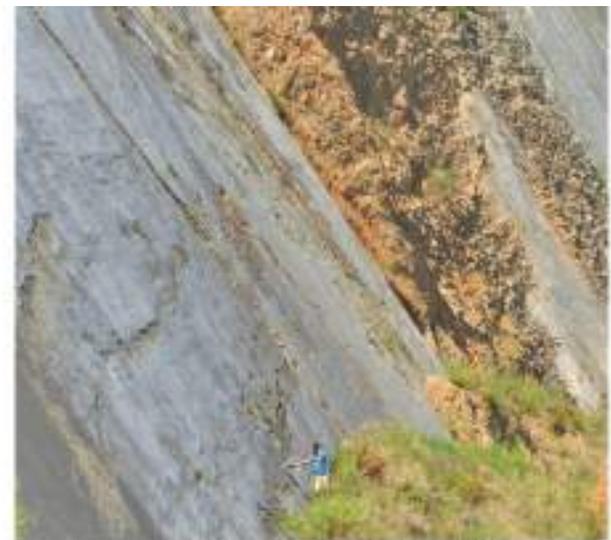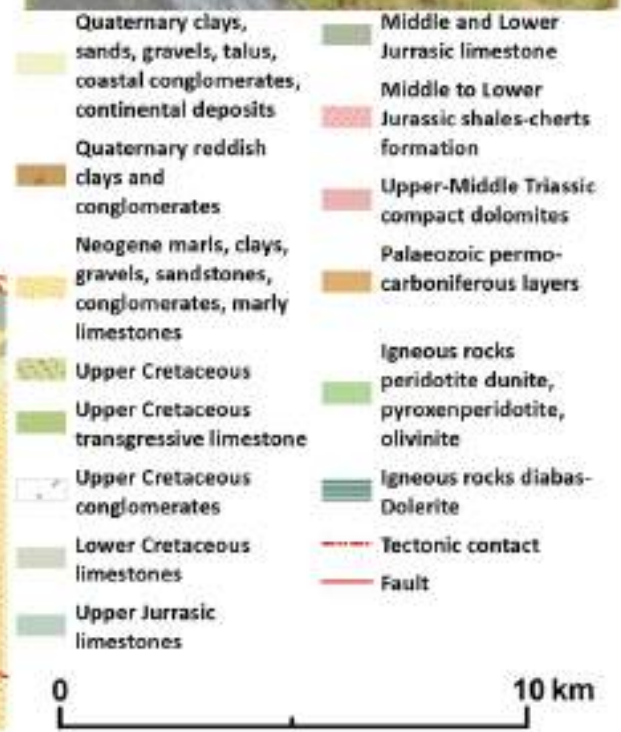

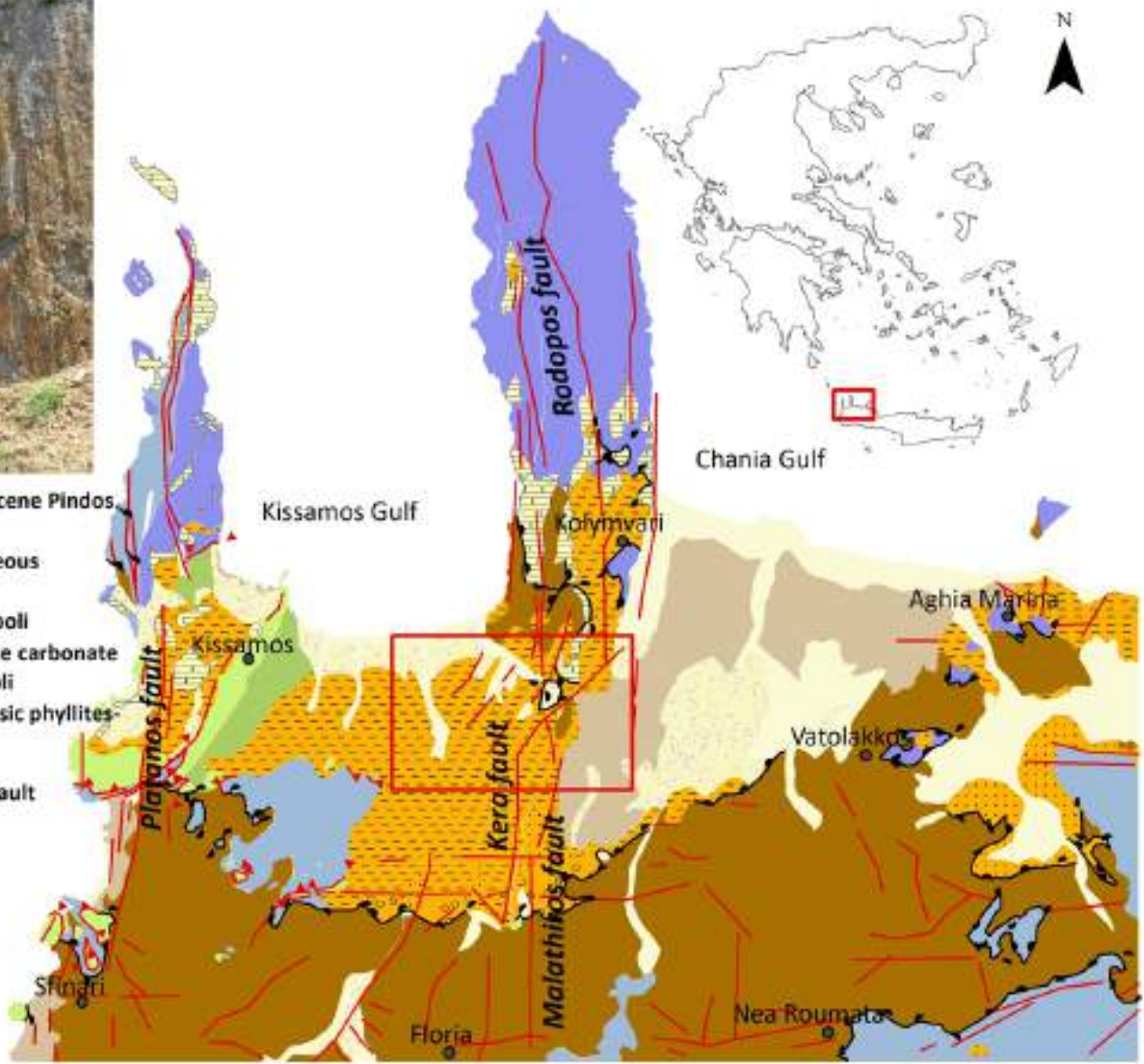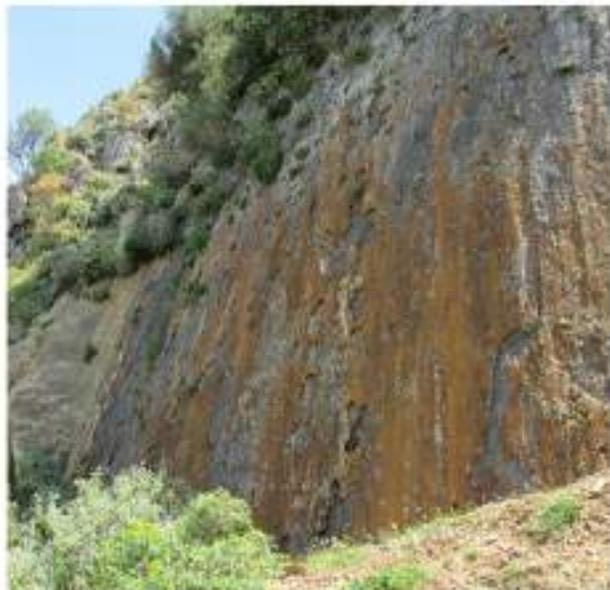

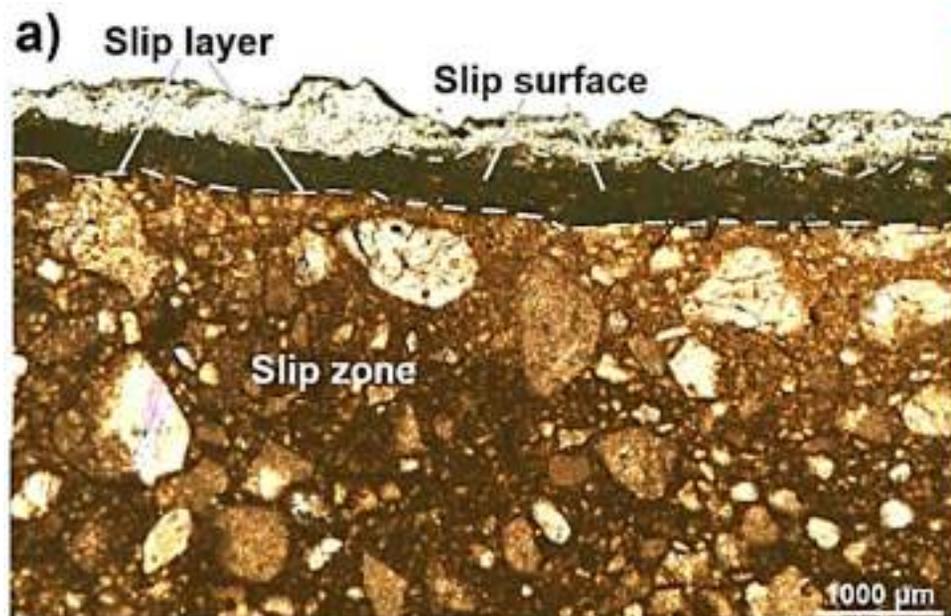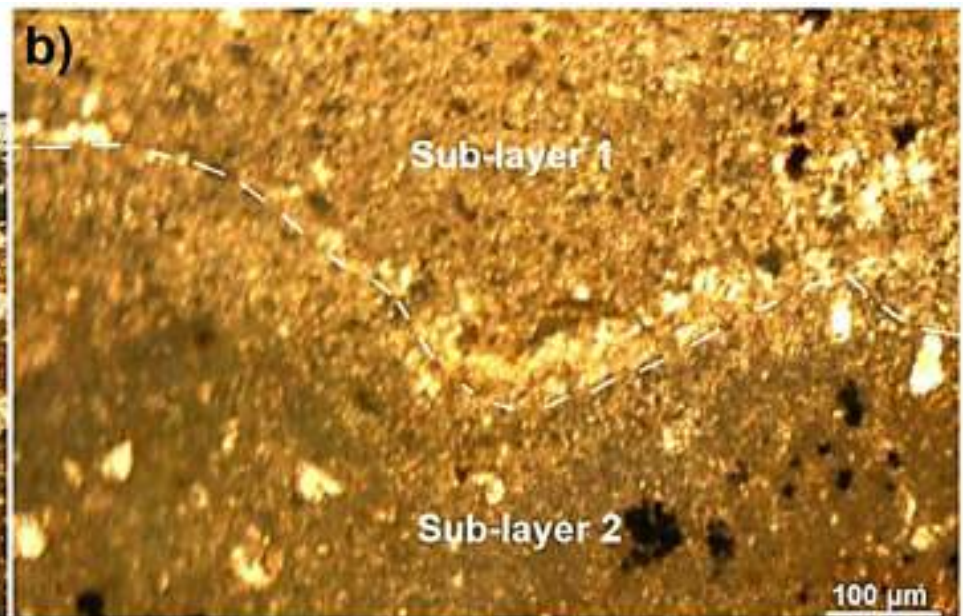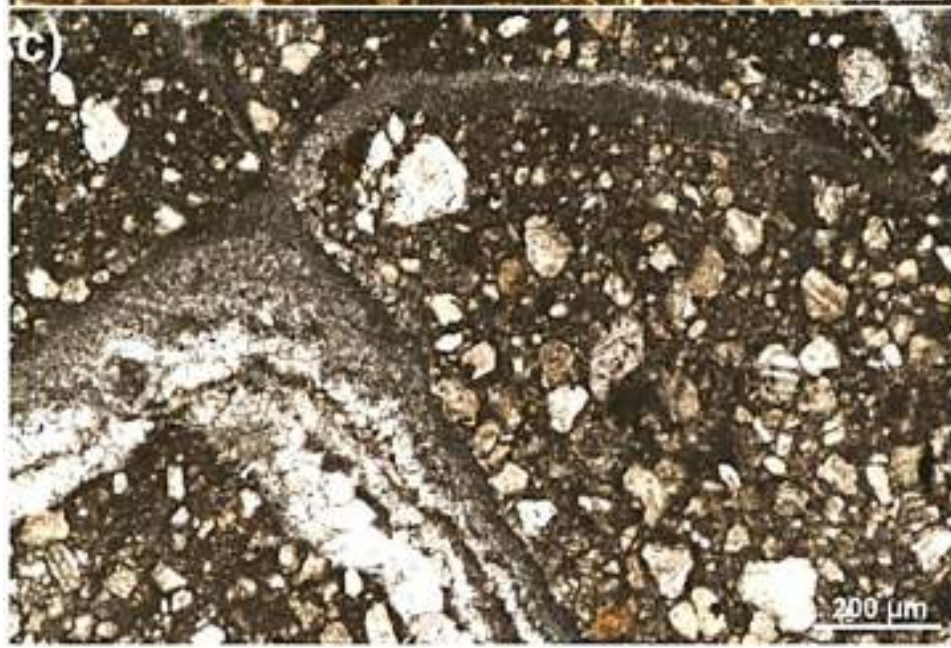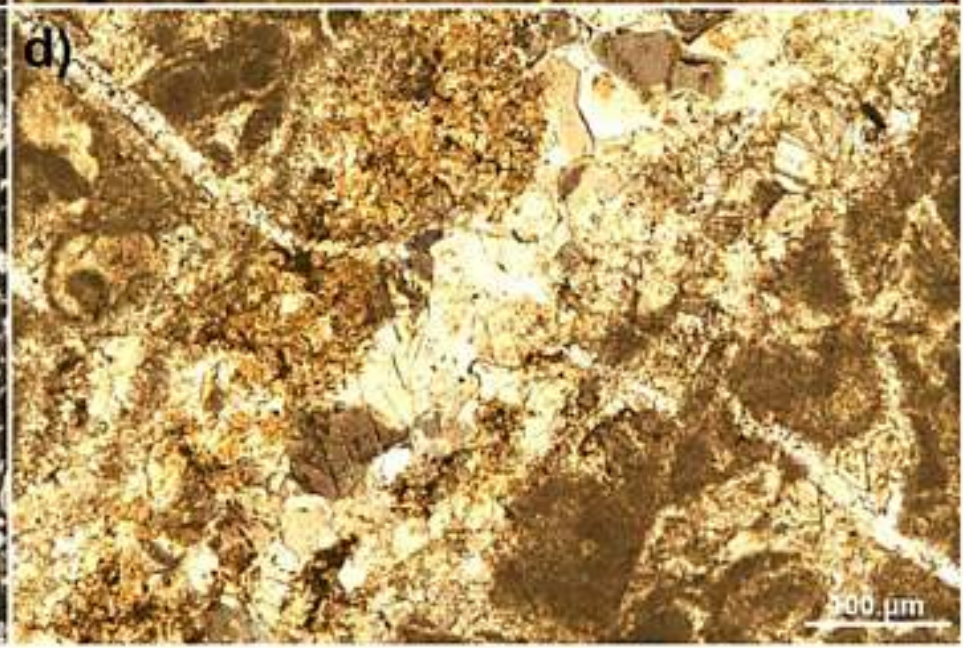

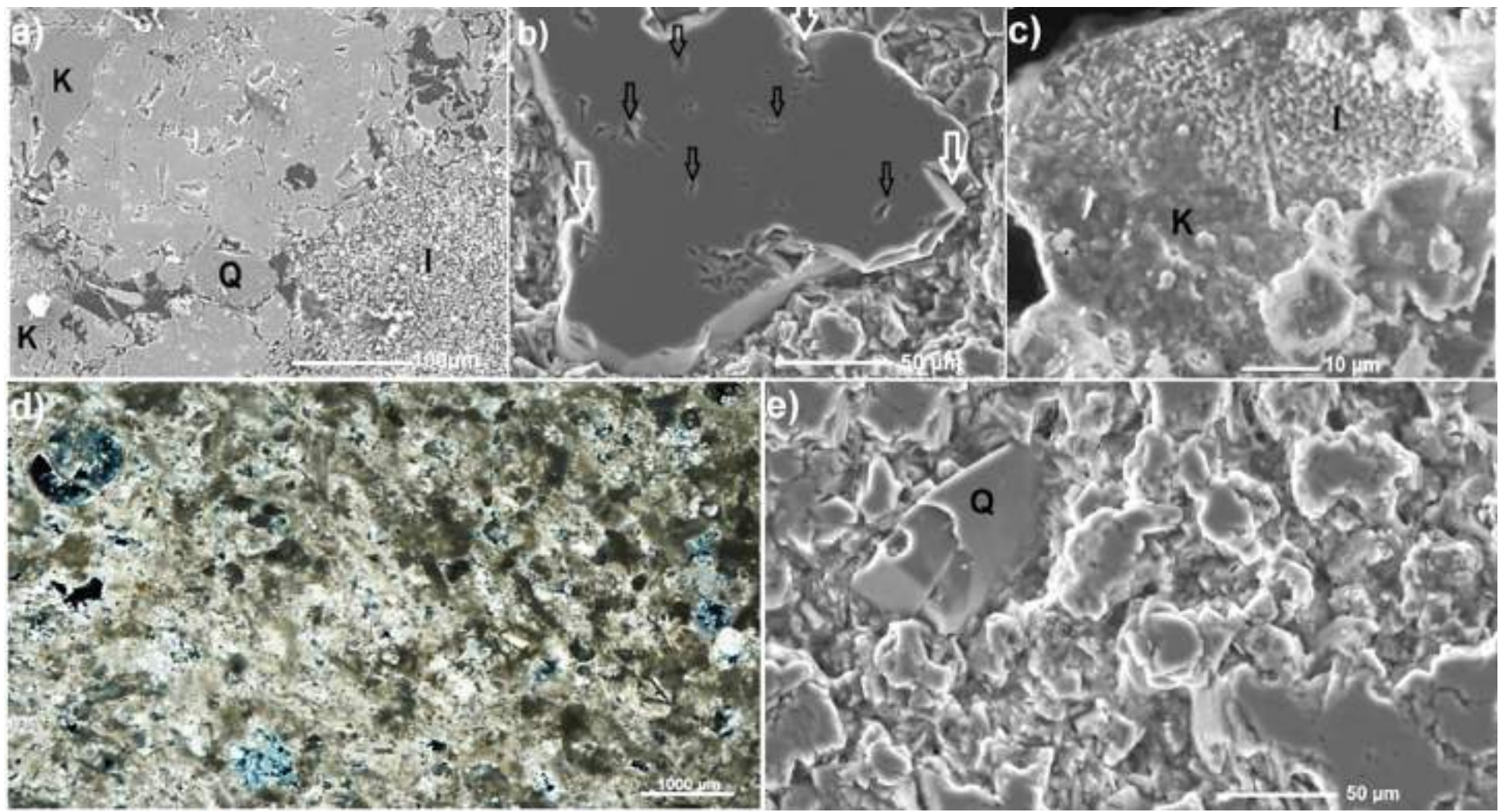

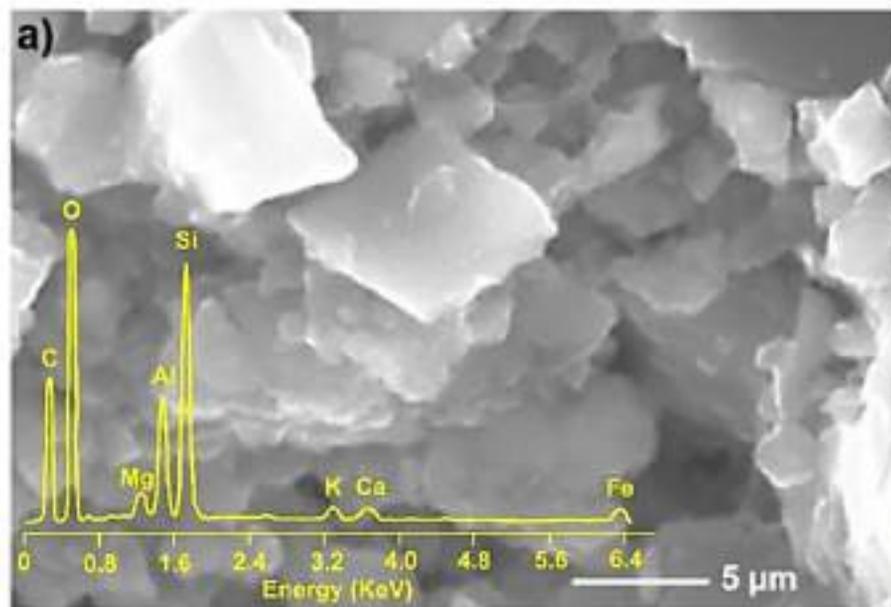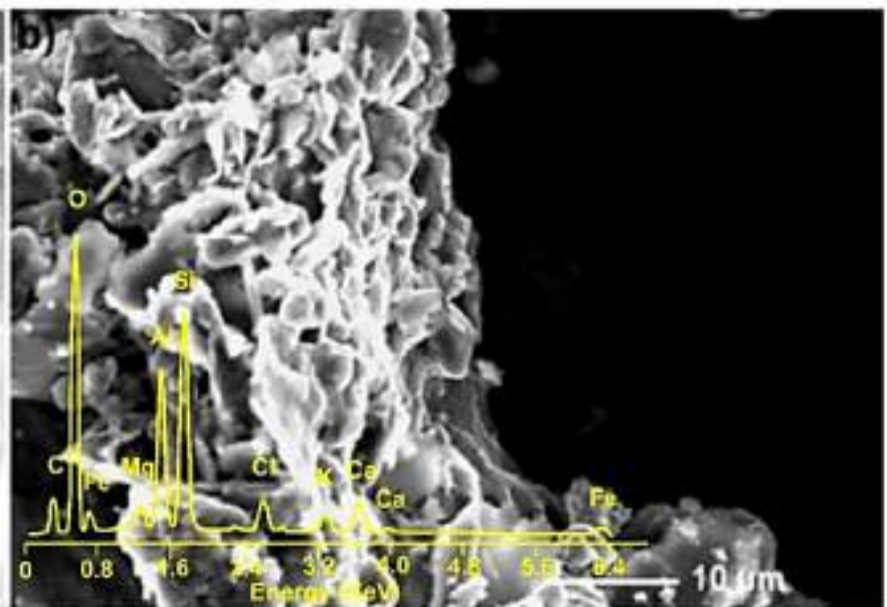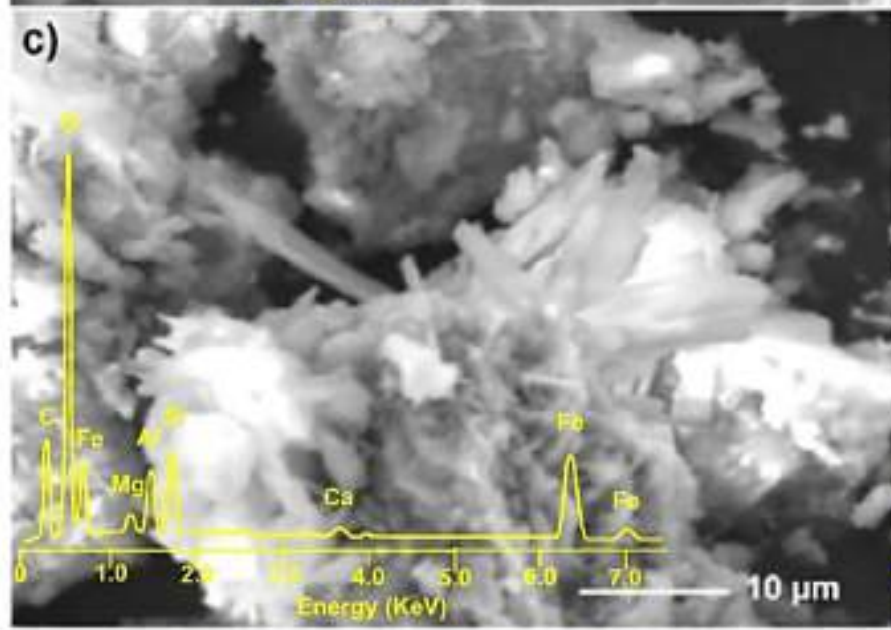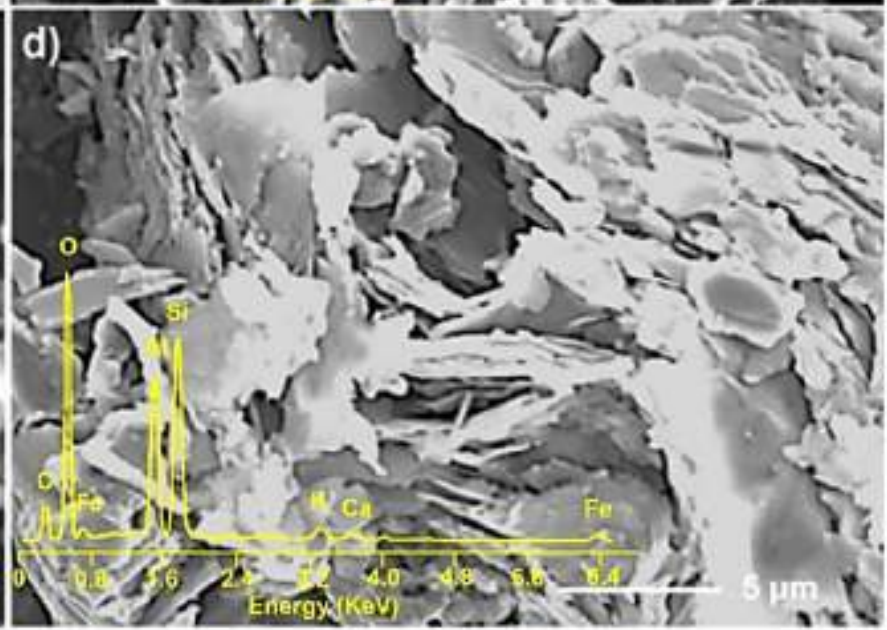

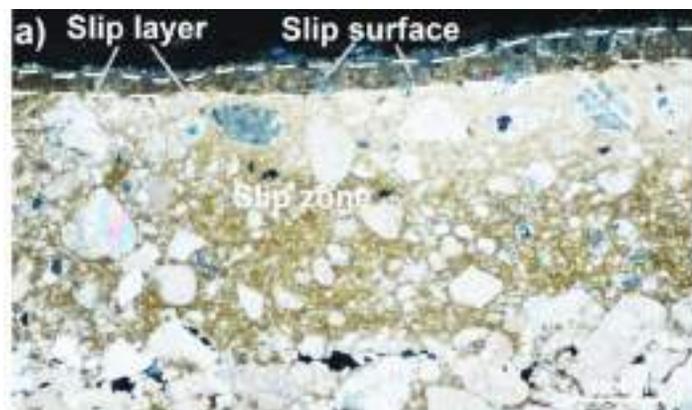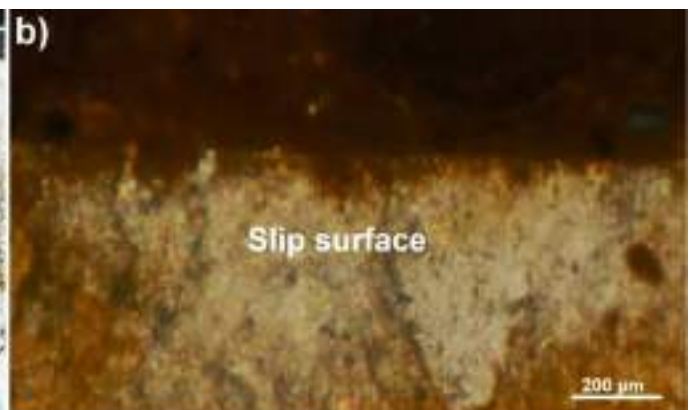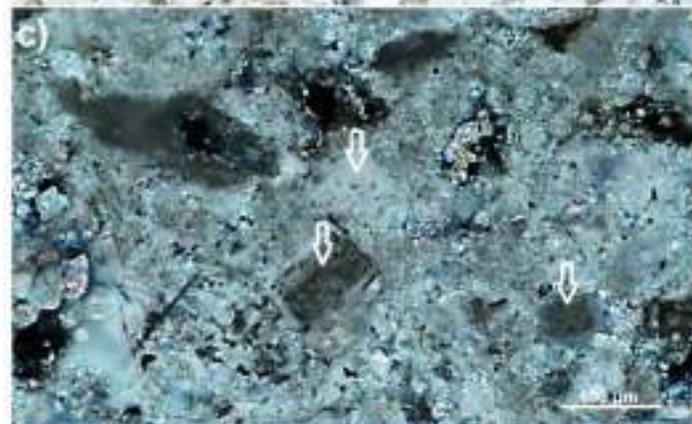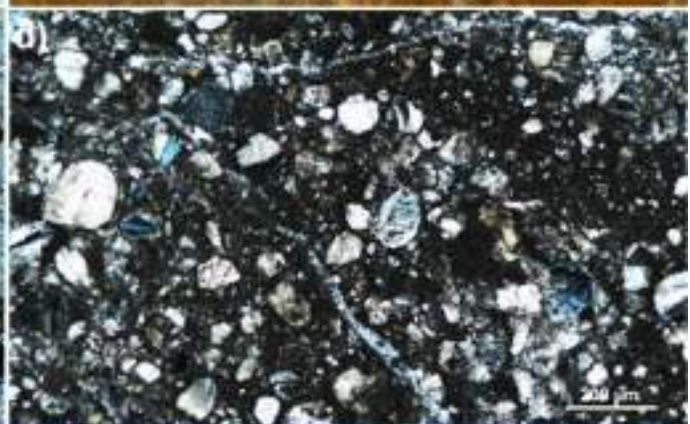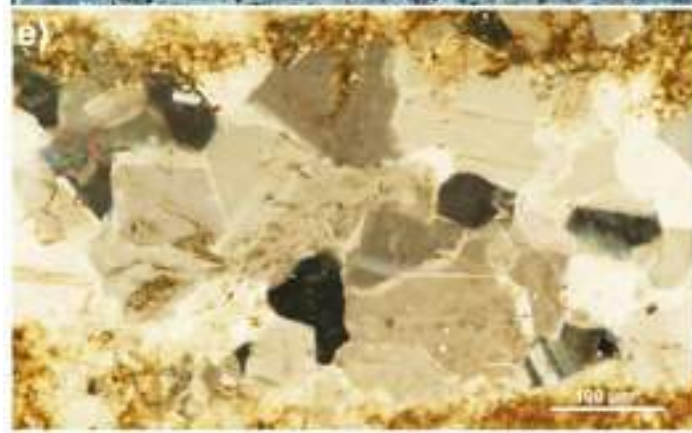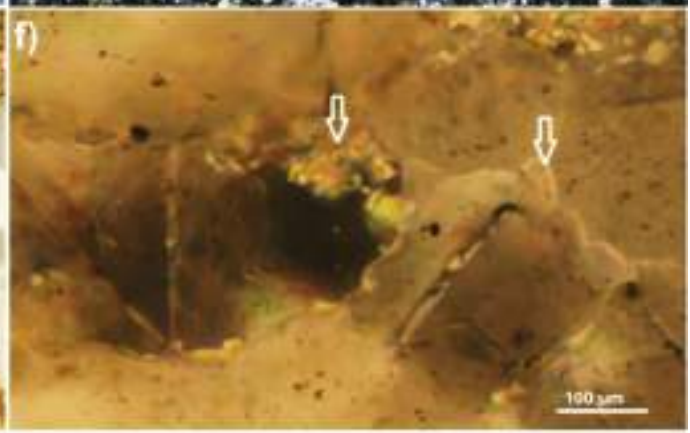

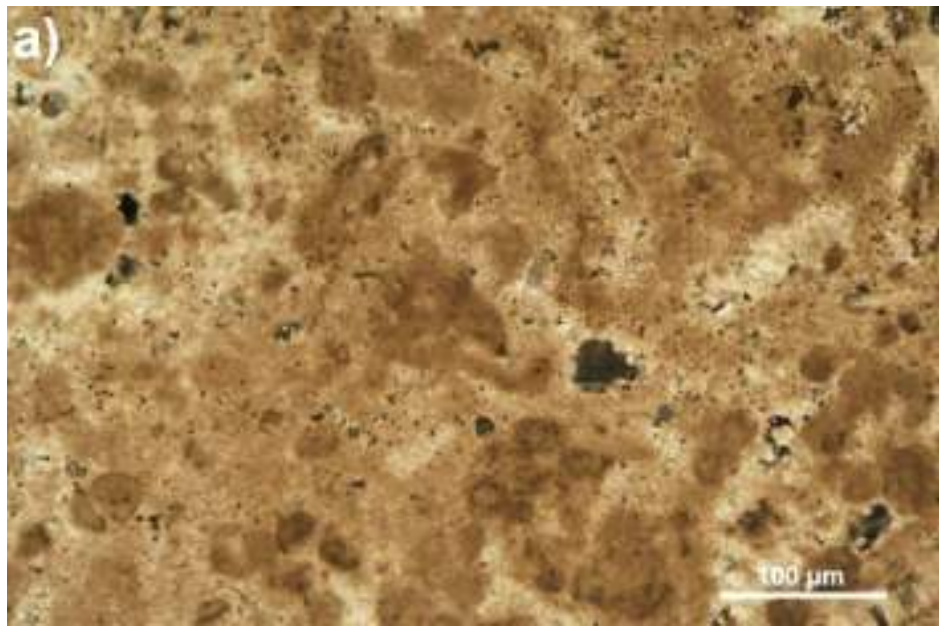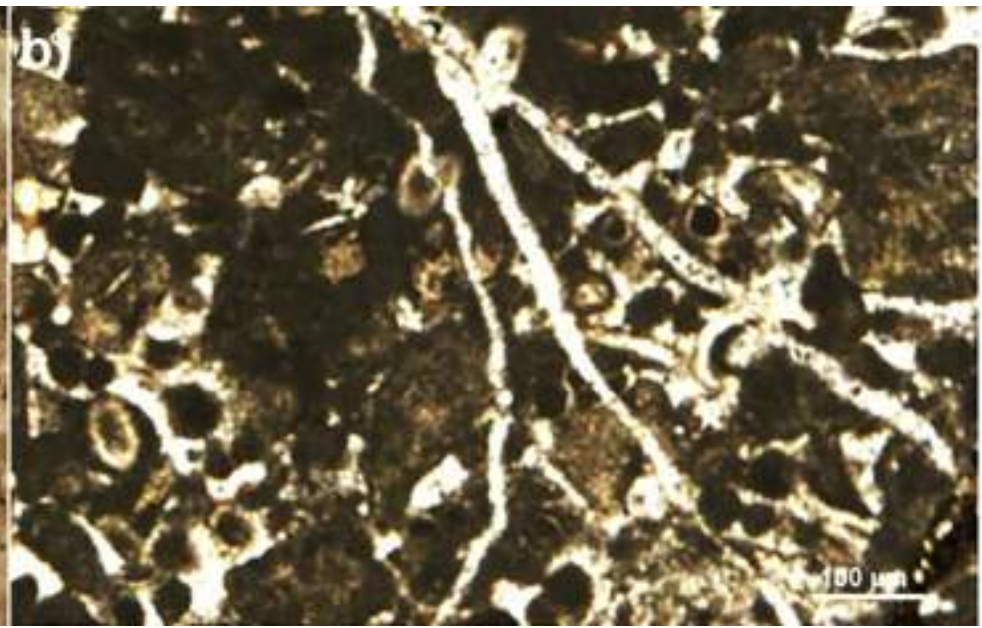

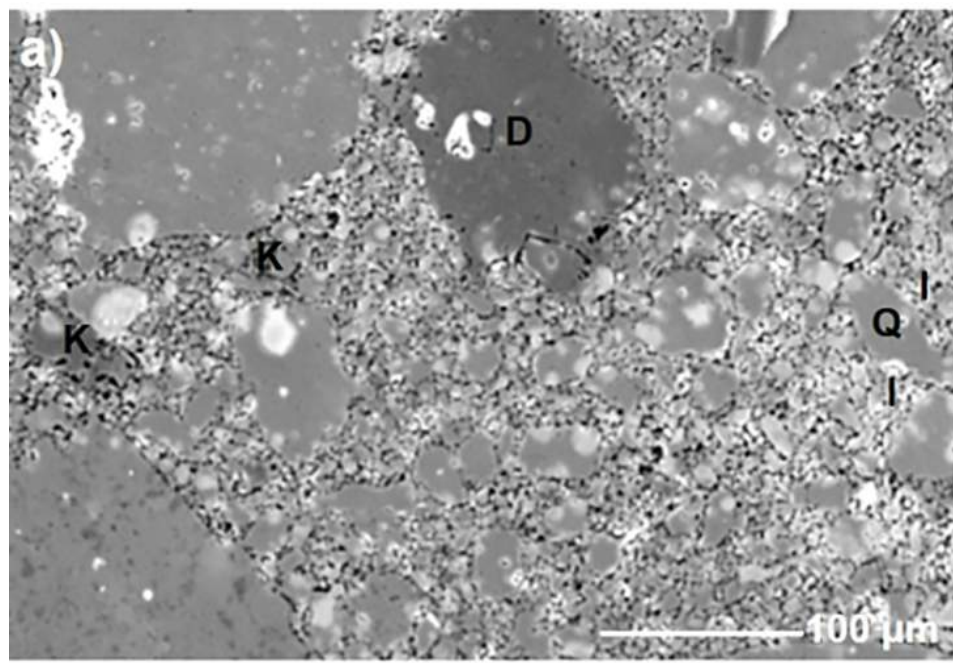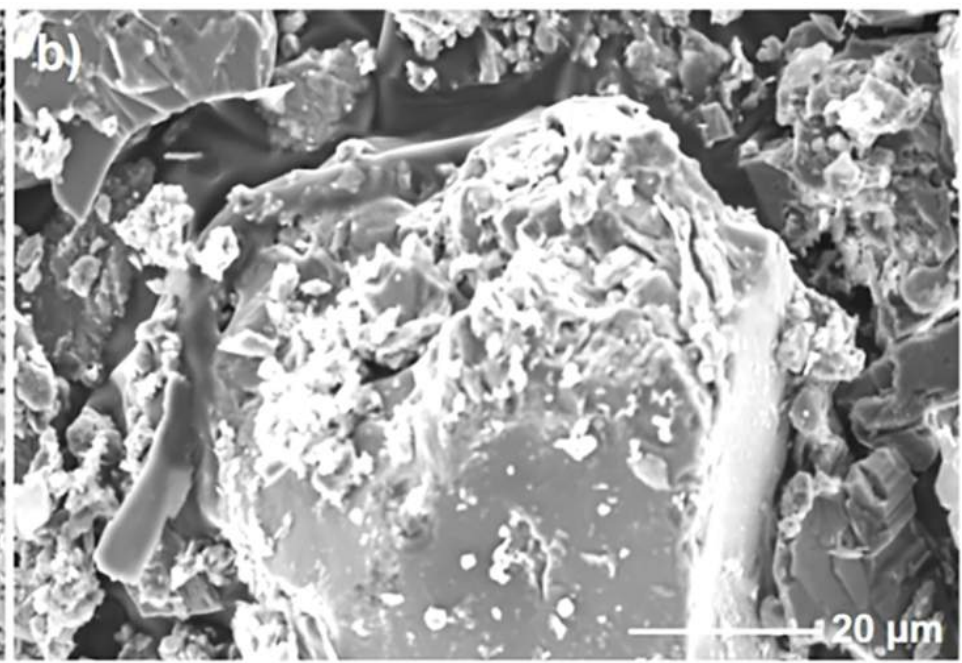

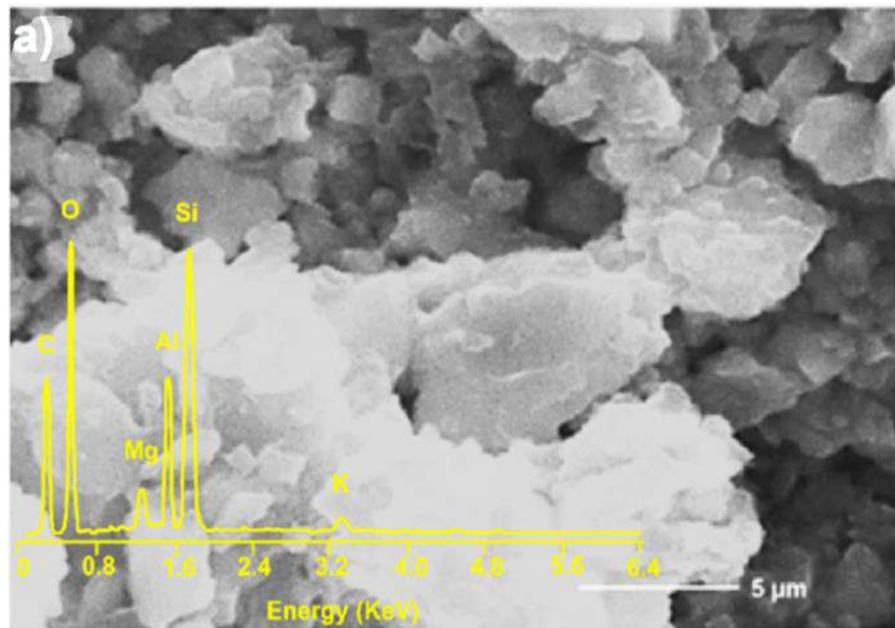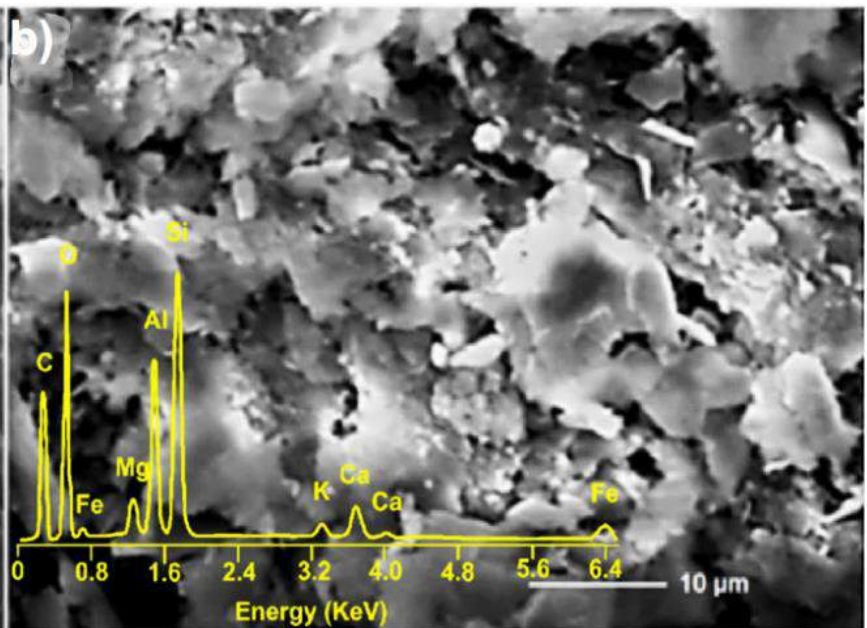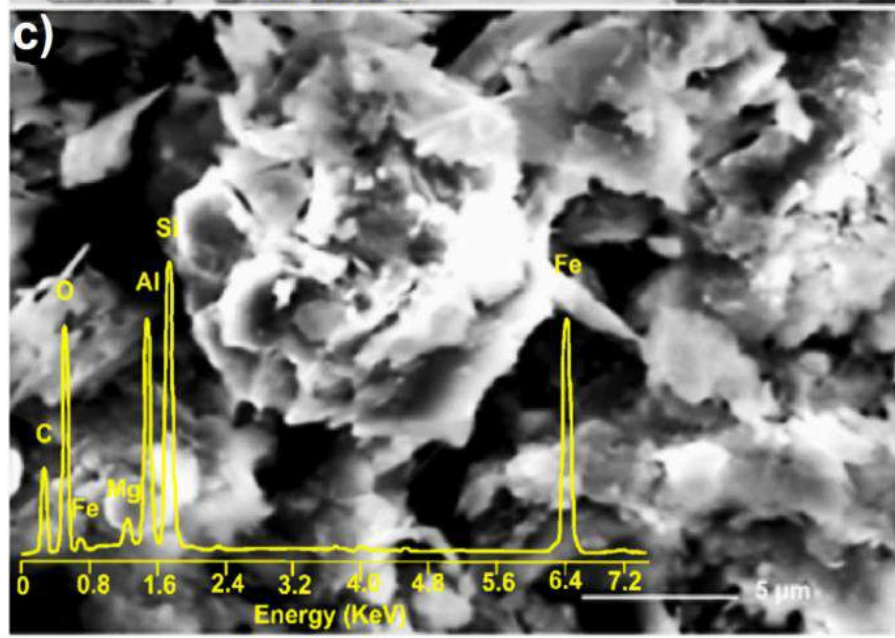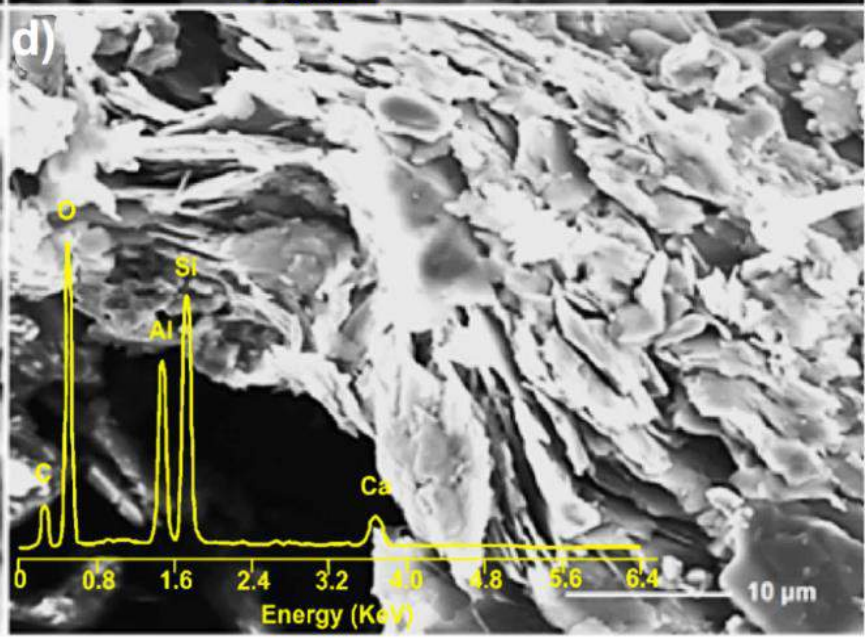

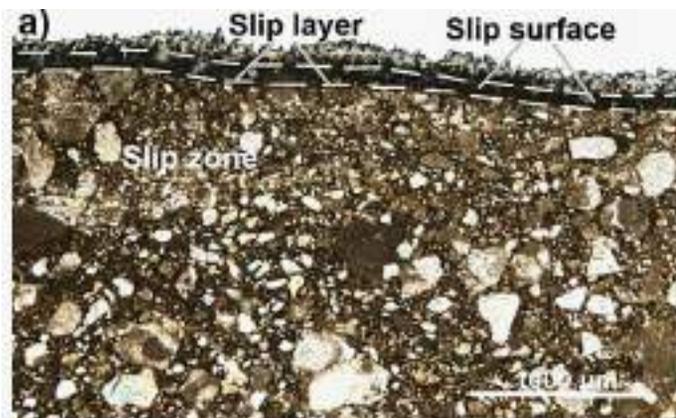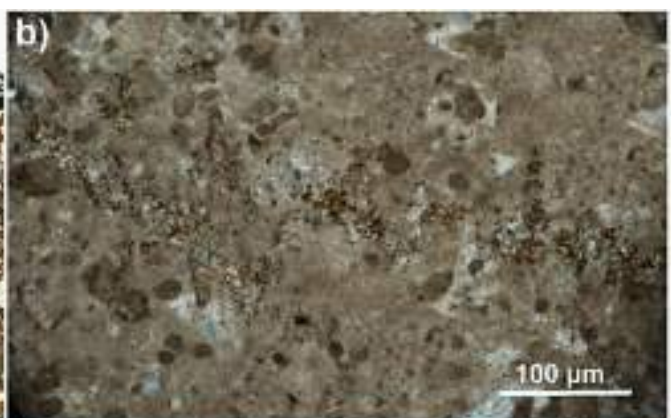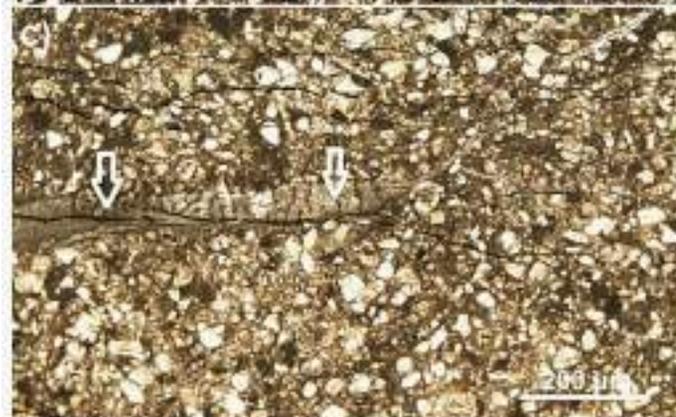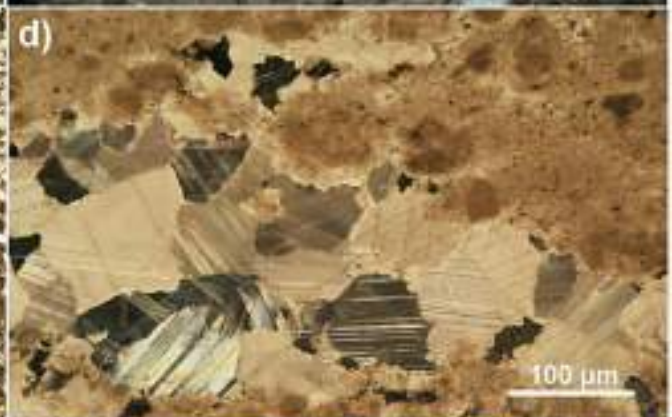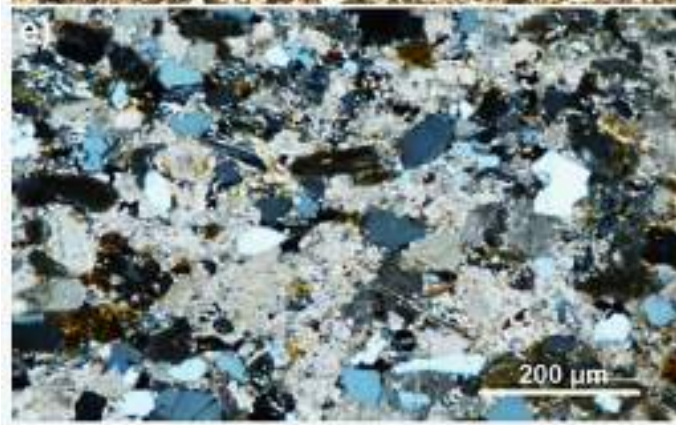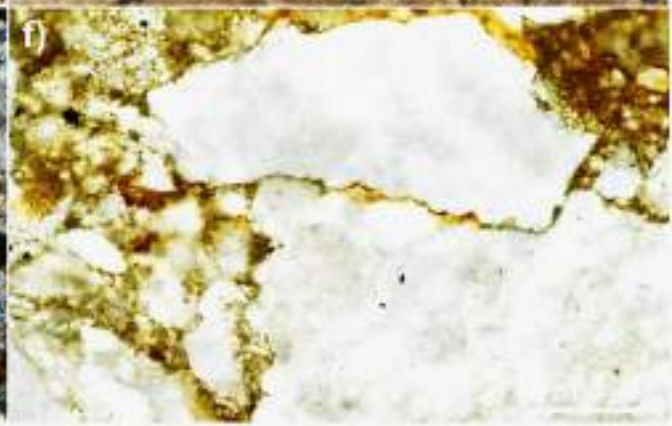

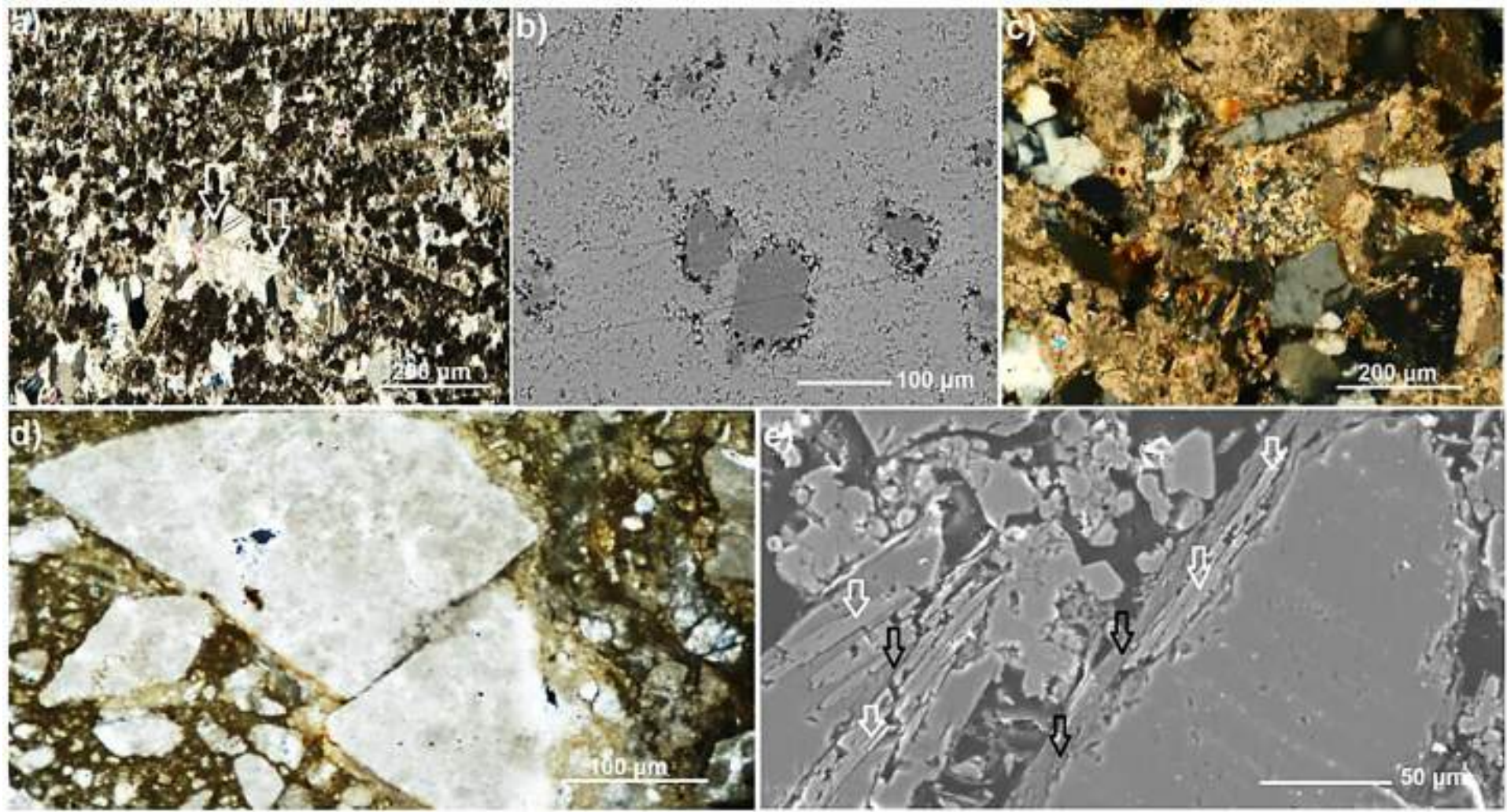

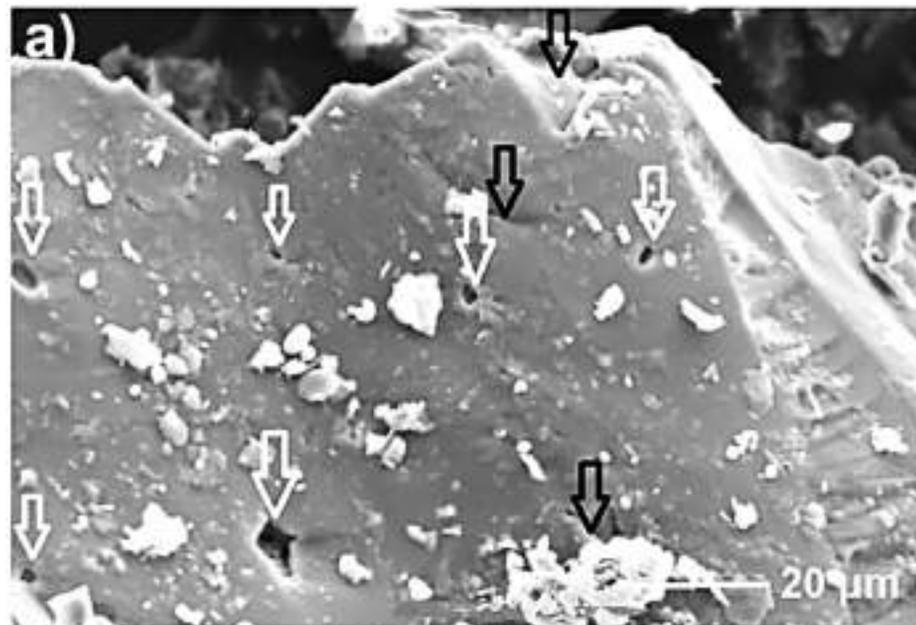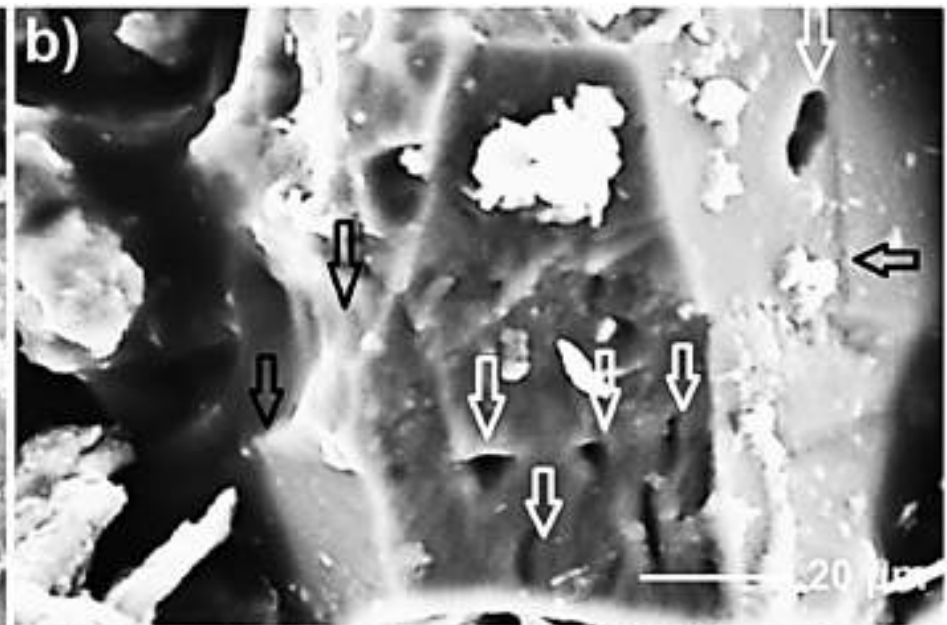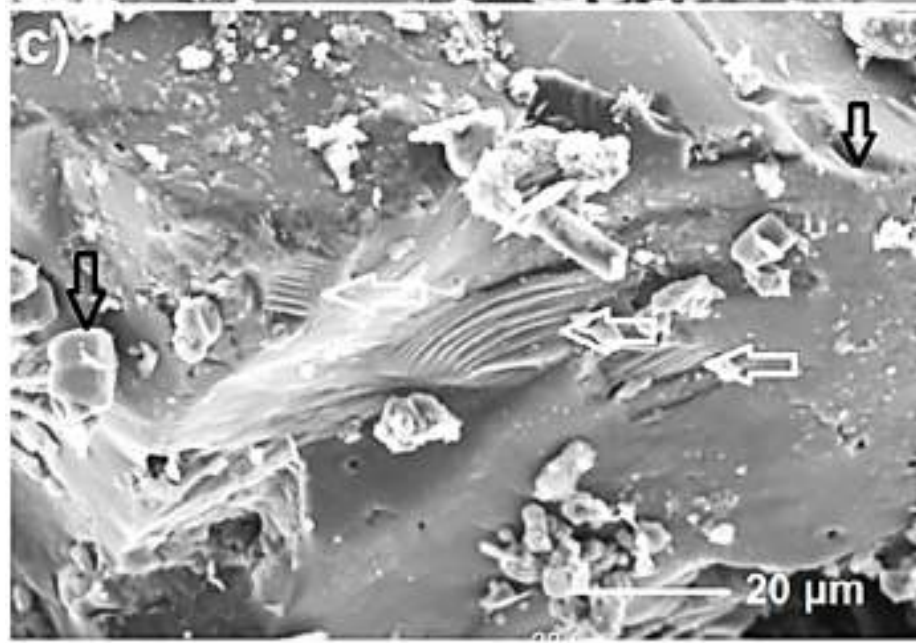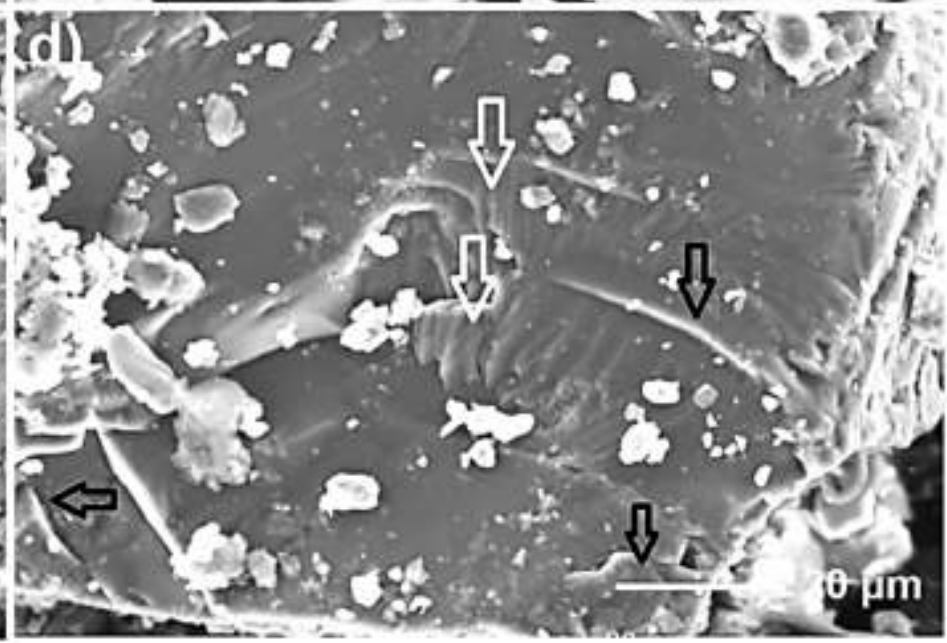

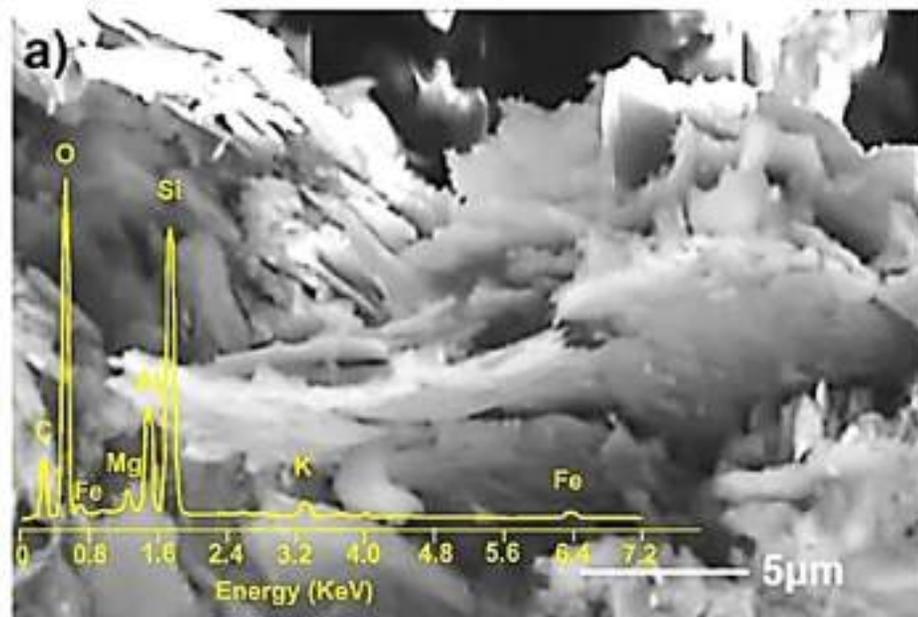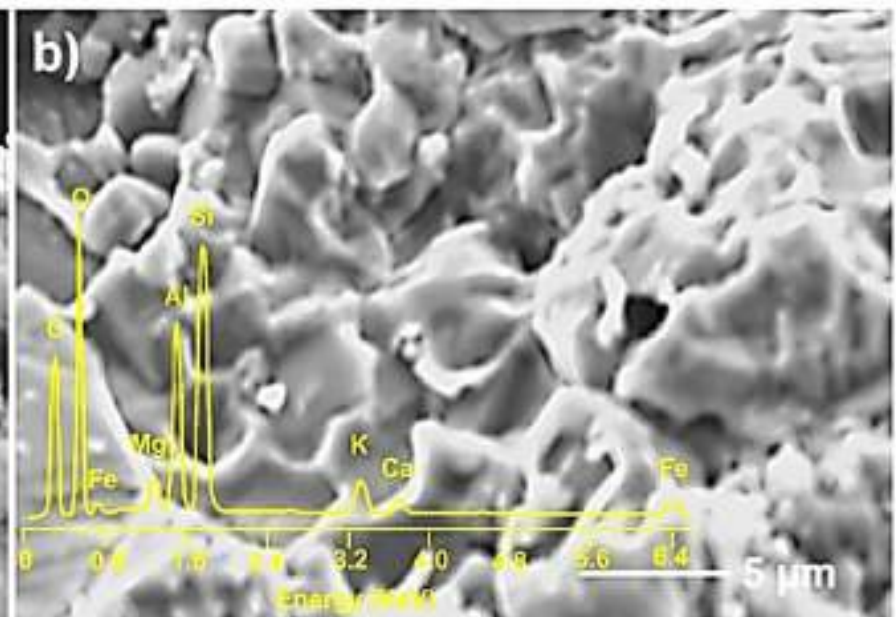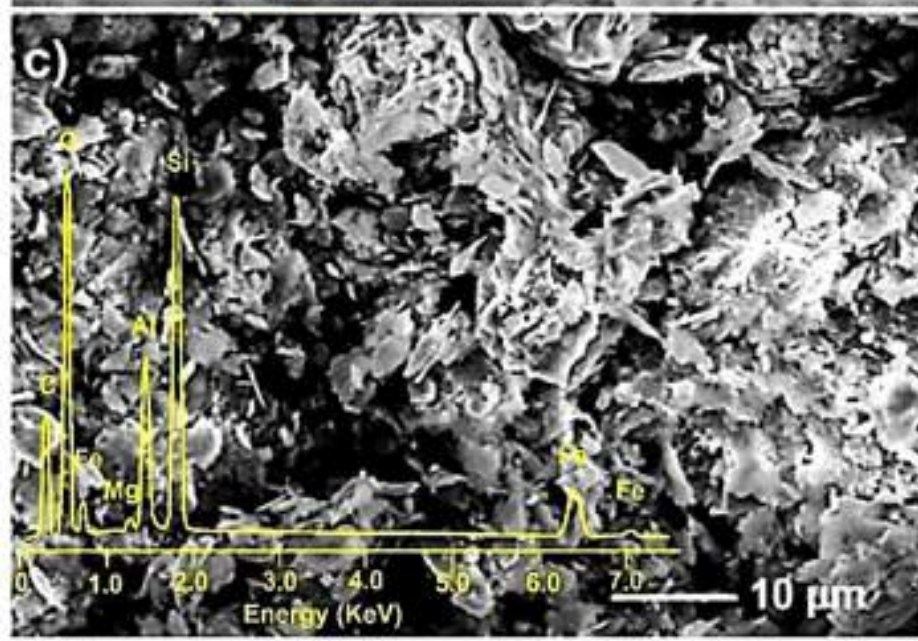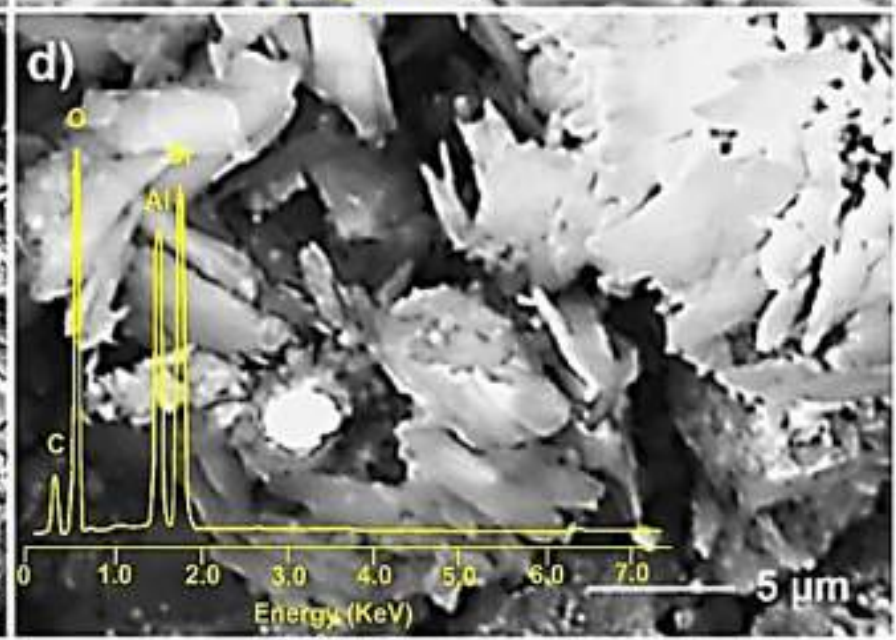